\documentclass[12pt]{article}
\usepackage[utf8]{inputenc}
\usepackage{amssymb,amsmath,color}
\usepackage{authblk}
\usepackage{appendix}
\usepackage{graphicx}
\usepackage{epstopdf}
\usepackage[sorting=none, backend=bibtex]{biblatex}
\usepackage{diagbox}
\usepackage[margin=1.0in]{geometry}
\bibliography{references}

\def \al {\alpha}

\def \om {\omega}

\def \ep {\epsilon}

\def \Jc {\mathcal{J}}

\def \Cc {\mathcal{C}}

\def \Oc {\mathcal{O}}

\def \pr {\partial}
\def \ra {\rightarrow}

\def \oh {\cfrac{1}{2}}

\def \beq { \begin{equation}}
\def \eeq {\end{equation}}

\renewcommand\Im{\operatorname{Im}}
\newcommand\const{\operatorname{const}}

\newcommand\Pf{\operatorname{Pf}}
\newcommand\Sch{\operatorname{Sch}}

\newcommand\EllipticK{\operatorname{EllipticK}}
\newcommand\EllipticF{\operatorname{EllipticF}}

\newcommand\sgn{\operatorname{sgn}}

\def \l {\left(}
\def \r {\right)}

\def \bra {\langle}
\def \ket {\rangle}

\def \AM#1 {{\color{blue} AM:#1 }}

\title{Universal Constraints on Energy Flow and SYK Thermalization}

\author[1]{Ahmed Almheiri \thanks{almheiri@ias.edu}}
\author[2]{Alexey Milekhin \thanks{milekhin@princeton.edu}}
\author[3]{Brian Swingle \thanks{bswingle@umd.edu}}

\affil[1]{Institute for Advanced Study, Princeton, NJ}
\affil[2]{Physics Department, Princeton University, Princeton, NJ}
\affil[3]{Condensed Matter Theory Center, Maryland Center for Fundamental Physics, Joint Institute for Quantum Information and Computer Science, and Department of Physics, University of Maryland, College Park}

\date{}

\begin{document}

\maketitle

\begin{abstract}
    We study the dynamics of a quantum system in thermal equilibrium that is suddenly coupled to a bath at a different temperature, a situation inspired by a particular black hole evaporation protocol. We prove a universal positivity bound on the integrated rate of change of the system energy which holds perturbatively in the system-bath coupling. Applied to holographic systems, this bound implies a particular instance of the averaged null energy condition. We also study in detail the particular case of two coupled SYK models in the limit of many fermions using the Schwinger-Keldysh non-equilibrium formalism. We solve the resulting Kadanoff-Baym equations both numerically and analytically in various limits. In particular, by going to low temperature, this setup enables a detailed study of the evaporation of black holes in JT gravity.
\end{abstract}

\newpage

\tableofcontents

\newpage

\section{Introduction}

Motivated by numerous recent experiments probing the out-of-equilibrium dynamics of reasonably well isolated quantum many-body systems, e.g.~\cite{Kaufman_2016,Bernien_2017,Zhang_2017}, and by long-standing theoretical questions concerning the nature of information processing in complex quantum systems, there has been a recent surge of interest in the physics of thermalization. In general, the phenomenon of thermalization is complex, involving many physical processes, including local relaxation of disturbances, diffusion of charge and energy, global spreading or scrambling of quantum information~\cite{Hayden:2007cs,Sekino_2008,brown2012scrambling}, and much more. This makes the subject complicated and rich.

Given this complexity, one natural starting point is to search for fundamental bounds on quantum dynamics. For example, in the context of strongly interacting many-body systems, physicists have speculated about a `Planckian' limit to scattering that might shed light on various material properties, e.g.~\cite{sachdev2001quantum,Kovtun_2005,Hartnoll_2014,Bruin804,hussey2004universality,PhysRevLett.122.216601,Legros_2018,Zaanen_2019}.\footnote{`Planckian' because the scattering time estimate uses only Planck's constant and the thermal scale: $ \frac{\hbar}{k_B T}$} In the context of quantum chaos, a similar kind of Planckian bound has been derived for the growth as chaos as diagnosed by so-called out-of-time-order correlators~\cite{Maldacena_2016}. One may wonder if such Planckian bounds can also be found for other aspects of thermalization.

In addition to general bounds, simple solvable models provide another powerful approach to understand quantum thermalization. Whereas bounds control the shape of the space of possibilities, solvable models give us archetypal behaviors or fixed points to which general models can be compared. In this context, considerable recent attention has been paid to the Sachdev-Ye-Kitaev (SYK) model~\cite{Sachdev1993Gapless,Georges2000Mean,Georges2001Quantum,kitaev,Polchinski_2016,ms,kamenev,Garc_a_Garc_a_2016} and its variants~\cite{Gu_2017,alex2019syk,Guo_2019,Can_2019a,Kim_2019,Can_2019b,dai2018global,Chowdhury_2018,Ben_Zion_2018} as tractable models of chaotic, thermalizing systems.

In this work, we study the equilibration of a system suddenly coupled to a large bath. The key object in our analysis is the energy curve: the time-dependence of the system energy after the system-bath coupling is suddenly turned on at zero time. At a schematic level, our results are as follows. First, we show that the energy curve has generic early time feature in which the system energy first increases with time even when the bath is cooler than the system. This initial increase is shown to obey a universal Planckian bound which constrains the shape of the early time energy bump. Second, we setup and analyze in detail a simple model of system-bath thermalization in which both system and bath are SYK models and the bath size is much greater than the system size. We are able to numerically compute the energy curve in this setup, including the early time energy rise and subsequent crossover to energy loss, the intermediate time draining of energy from the system, and the late time approach to equilibrium. In the low temperature limit, we also derive various analytical results, for example, the case of energy loss into a zero-temperature bath (Sec. \ref{sec:total_evaporation}).

Our results are related to the physics of black hole evaporation in AdS \cite{Rocha:2010zz,Engelsoy:2016xyb, Almheiri:2018xdw}. One precise connection can be made via the SYK model, which at low temperatures exhibits a dynamical sector that is identical to a form a quantum gravity in a two-dimensional nearly AdS spacetime. In this context, we show that our universal bound on the early time bump in the energy curve is equivalent to an instance of the averaged null energy condition. The latter is an important constraint on energy flow that is often assumed in general relativity. Moreover, our coupled SYK system-bath setup reproduces and generalizes simple phenomenological models of black hole evaporation in which absorbing boundary conditions were used to extract the radiation.

The rest of the paper is structured as follows. In the remainder of the Introduction we summarize our results in more detail. In Section 2 we setup and prove a rigorous bound on early time energy dynamics and demonstrate its relation to the averaged null energy condition in quantum gravity. In Section 3 we setup the coupled SYK cluster model. We analyse its equilibrium properties and use a Schwinger-Keldysh approach to analyse the system out-of-equilibrium. We report both numerical studies as well as comprehensive analytical results in various limits. In particular, Section \ref{sec:total_evaporation} is dedicated to studying total evaporation into a zero-temperature bath. The description of the exact numerical setup and detailed calculations can be found in Appendices. Section 4 contains a brief discussion of our results and possible future directions.

\subsection{Summary of results}

This section describes the setting for our results and summarizes again the main points in more technical language. We consider the interaction of a system $S$ with a bath $B$ which is much larger than the system. This allows one to ignore the backreaction of the system on the bath. The system and bath have Hamiltonians $H_S$ and $H_B$, respectively, and at  time zero they are coupled via $g H_{SB}$. The goal is to understand how the system energy changes as a function of time due to this coupling.

Just before the coupling is turned on, the system and bath are in independent thermal states at inverse temperatures $\beta_{S}^{(0)}$ and $\beta_B^{(0)}$, respectively. The time evolution of the system-bath composite is
\begin{equation}
    \rho_{SB}(t) = e^{-i H t}\left(  \frac{e^{-\beta_S^{(0)}H_S}}{Z_S}\otimes \frac{e^{-\beta_B^{(0)}H_B}}{Z_B}\right) e^{i H t}
\end{equation}
where 
\begin{equation}
    H = H_S + H_B + g H_{SB}
\end{equation}
and
\begin{equation}
    H_{SB}=O_S O_B
\end{equation}
is a product of two Hermitian operators. We also present some numerical calculations (Section~\ref{sec:syk_ed}) where the system and bath are initialized into pure states.

\begin{figure}
    \centering
    \includegraphics[width=.8\textwidth]{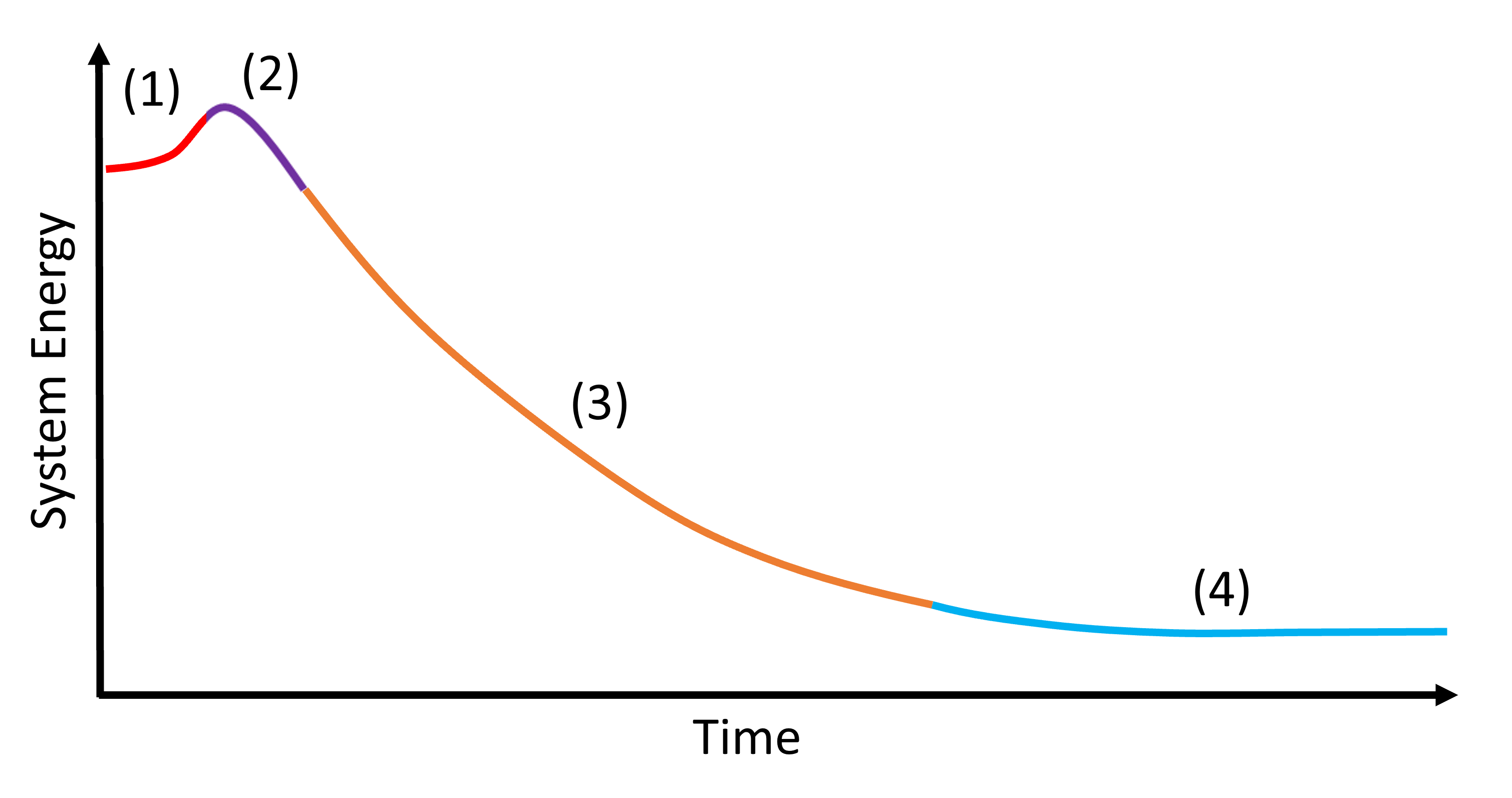}
    \caption{Typical behavior of system energy as a function of time for a large bath at lower temperature. We distinguish four dynamical regimes, labeled (1), (2), (3), and (4), which are discussed in detail in the text. Roughly they correspond to the early time energy rise, the subsequent turnover to energy loss, a sustained period of energy loss, and the final approach to global equilibrium.}
    \label{fig:energy_curve_sketch}
\end{figure}

The primary observable of interest is the energy curve of the system,
\begin{equation}
    E_S(t) = \text{tr}[ H_S \rho_{SB}(t)].
\end{equation}
A typical energy curve is sketched in Figure~\ref{fig:energy_curve_sketch}. Assuming the bath is cooler than the system, there are four key pieces of the energy curve: (1) the very early time energy increase, (2) the subsequent turnover to energy loss, (3) a sustained period of quasi-steady-state energy loss, and (4) a final approach to true system-bath equilibrium. 

The first main result is a general bound on the energy curve whenever the system-bath interaction is a single product of operators. For simplicity, consider a limit where the system-bath coupling $g$ is small, so that the system temperature is approximately constant on the time-scale of the inherent system dynamics. Define the integrated energy flux by
\begin{equation}
\label{the_bound}
    F_\kappa = \int_0^\infty dt e^{- \kappa t} \dot{E}_S.
\end{equation}
We show that this quantity is guaranteed to be positive for sufficiently large $\kappa$:
\begin{equation}
    \kappa \geq 2/\beta_S \implies F_\kappa \geq 0 .
\end{equation}
This result is proven for any system and any bath to leading order in perturbation theory in $g$. In the context of SYK, we show that it holds more generally (Section~\ref{sec:syk_bound}). The constant $\kappa$ sets the time-scale; reintroducing Planck's constant $\hbar$ and Boltzmann's constant $k_B$, the boundary value of $\kappa$ is
\beq
\frac{2}{\beta_S} = \frac{2\hbar}{T_S k_B} = 2.6 \times 10^{-14}\ s \ {\rm at} \ T_S=293\ K 
\eeq

The other main results are obtained in a particular model in which both system and bath are SYK clusters. We consider two SYK models, a system composed of $N$ fermions with $q_S$-body interactions and a bath composed of $M$ fermions with $q_B$-body interactions. The system and bath are coupled via a random term involving $f_S$ system fermions and $f_B$ bath fermions. We take $M \gg N$ so that the bath is unaffected by the coupling to the system. See Section~\ref{sec:syk} for more details and Ref.~\cite{Chen2017} for another study of two coupled SYK clusters. 

We derive the full large $N$, large $M$ Schwinger-Keldysh equations of motion for this system and numerically solve them following the technique in Refs.~\cite{quench1,quench2}. This allows us to compute the entire energy curve for this system-bath model as a function of the initial system temperature, the initial bath temperature, and all the other parameters of the model. 

Moreover, in the the low temperature limit we are able to solve the Kadanoff-Baym equation analytically to determine properties of the initial energy bump, the rate of energy loss, and the approach to final equilibrium.

Finally, using the gravitational description of the low energy dynamics of SYK, we argue that our general bound on energy flux is equivalent to one instance of a bulk energy bound called the average null energy condition. Specifically, we show that the positivity of the energy flux for $\kappa \geq 2\pi/\beta_S$ implies the ANEC in the bulk integrated over the black hole horizon. Curiously, the condition $\kappa \geq 2 \pi/\beta_S$ is actually weaker than the most general condition proven in perturbation theory, which is $F_\kappa \geq 0$ for all $\kappa \geq 2/\beta_S$.

Note: While this paper was in the final stages of preparation, Ref.~\cite{zhang2019evaporation} appeared on the arXiv which has some overlap with our results.

\section{Bounds on energy dynamics}
\label{sec:bound}

In this section we discuss the general positivity bound on the integrated energy flux introduced above. This bound holds perturbatively in the system-bath coupling whenever the system-bath interaction is a simple product form, $H_{SB} = O_{S} O_B$. In subsequent subsections, we discuss the general situation with multiple operator couplings and the relation to energy conditions in holography.

\subsection{Perturbative bound}

Recall that the integrated energy flux is defined by
\begin{equation}
    F_\kappa = \int_0^\infty dt e^{- \kappa t} \dot{E}_S.
\end{equation}
In Appendix~\ref{app:bound} we prove that 
\begin{equation}
    F_\kappa \geq 0 \,\,\, \text{for}\,\,\, \kappa \geq \frac{2}{\beta_S}
\end{equation}
in the weak coupling limit, $g \rightarrow 0$, for any system and bath Hamiltonians.

The proof proceeds by explicitly computing the integrated flux $F_\kappa$ in terms of spectral functions associated with the system operator $O_S$ and the bath operator $O_B$. The positivity of the spectral functions can then be used to constrain the integrated flux. Making no other assumptions about the system and bath spectral functions, one can show that $F_\kappa \geq 0$ for all $\kappa \geq 2/\beta_S$. With further assumptions on the system or bath, it might be possible to strengthen this result.

The details are in Appendix~\ref{app:bound}, but a few key formulas are reproduced here. To begin, we define the spectral function for an operator $O$ via the response function,
\begin{equation}
    X^R(t) = -i \theta(t) \langle [O(t),O(0)]\rangle_\beta.
\end{equation}
The Fourier transform is denoted $X^R(\omega)$, and the spectral function $A(\nu)$ is then
\begin{equation}
    X^R(\omega) = \int \frac{d\nu}{2\pi} \frac{A(\nu)}{\omega+i0^+ -\nu}.
\end{equation}
We may further decompose the spectral function $A(\nu)$ into two positive definite pieces,
\begin{equation}
    A(\nu) = A_+(\nu) - A_-(\nu),
\end{equation}
defined by
\begin{equation}
    A_\pm(\nu) = 2\pi \sum_{n,m} p_n |\langle n |O |m \rangle |^2 \delta(\nu \mp [E_m - E_n])
\end{equation}
where $p_n = e^{-\beta E_n}/Z$ is the thermal probability.

The integrated flux in terms of the spectral functions $A_{S+}$ and $A_{B+}$ is
\begin{equation}
  F_\kappa =   2 g^2 \int \frac{d\omega}{2\pi}\frac{d\omega'}{2\pi} \frac{\omega A_{S+}(\omega) A_{B+}(\omega')}{(\omega+\omega')^2 + \kappa^2}.
\end{equation}

The short-time limit, corresponding the initial rise of energy [part (1) of Figure~\ref{fig:energy_curve_sketch}], can be accessed by taking $\kappa \rightarrow \infty$ to give
\begin{equation}
    F_{\kappa \rightarrow \infty} \rightarrow 2 \frac{g^2}{\kappa^2}  \int \frac{d\omega}{2\pi}\frac{d\omega'}{2\pi} \omega A_{S+}(\omega) A_{B+}(\omega').
\end{equation}
Using 
\begin{equation}
    \int d\omega \omega A_+(\omega) = \frac{1}{2} \int d\omega \omega A(\omega) = \int_{\omega \geq 0} d\omega \omega A(\omega) \geq 0,
\end{equation}
it follows that 
\begin{equation}
    F_{\kappa \rightarrow \infty} \geq 0
\end{equation}
in agreement with results in~\cite{Almheiri:2018xdw}.

The long-time limit, corresponding to the steady loss of energy [part (3) of Figure~\ref{fig:energy_curve_sketch}], can be accessed by taking $\kappa \rightarrow 0$ to give
\begin{equation}
    F_{\kappa \rightarrow 0} \rightarrow - \frac{ g^2}{\kappa}\int_{\omega \geq 0} \frac{d\omega}{2\pi} \omega \frac{\sinh \frac{(\beta_B - \beta_S)\omega}{2} A_S(\omega) A_B(\omega)}{2 \sinh \frac{\beta_S \omega}{2} \sinh \frac{\beta_B \omega}{2}}.
\end{equation}
This expression shows that energy always flows from hot to cold on these timescales. Note that we are not accessing the final approach to equilibrium since the coupling $g$ is being treated perturbatively and we are not yet studying the time-dependence of the system temperature.

\subsection{Multi-operator couplings}

It is natural to ask if the bound if the bound can be extended to include more general system-bath couplings. Consider a coupling of the form 
\begin{equation}
    H_{SB} = \sum_\alpha O_S^\alpha O_B^\alpha.
\end{equation}
In this case, a more general expression for the integrated flux can be derived which involves mixed correlators of $O^\alpha$ with $O^\beta$. We have not recorded this expression here because, as we show by example shortly, the integrated flux in this case does not obey a general positivity condition.\footnote{We thank Daniel Ranard for discussions on multi-operator couplings.}

If the correlations between $O^\alpha$ and $O^\beta$ are diagonal, then the positivity result continues to hold. This is because the diagonal terms reduce to the single product of operators case considered above. Although this is a special case, it is not an uncommon situation; for example, in the SYK model, different fermions are approximately decorrelated to leading order in $N$.

However, for a generic multi-operator coupling the energy may go down initially. For example, 
consider a single-qubit system $s$ interacting with a bath qubit $b$. The unperturbed Hamiltonian reads as:
\begin{equation}
    H_S+H_B = \omega_0 s^\dagger s + \Omega b^\dagger b 
\end{equation}
where $s$ and $b$ are lowering operators in a two-level system. We switch on the following quadratic interaction at time $t=0$:
\begin{equation}
    H_{SB} = V (s^\dagger b + b^\dagger s)
\end{equation}
It is easy to solve this quadratic theory exactly. If the initial density matrices are diagonal,
\begin{equation}
    \rho_{S(B)}(t=0) = \text{diag}(1-n_{S(B)}, n_{S(B)}), 
\end{equation}
then the expression for the system's energy at early times is given by:
\begin{equation}
    E_S(t) = \omega_0 \bra s^\dagger s \ket(t) = \omega_0 n_S + \omega_0 V^2 t^2(n_B - n_S) + \dots.
\end{equation}
From this expression it is obvious that the system energy may go down initially.

\subsection{Relation to energy conditions in holography}

\begin{figure}[t!]
\centering
\includegraphics[scale=0.75]{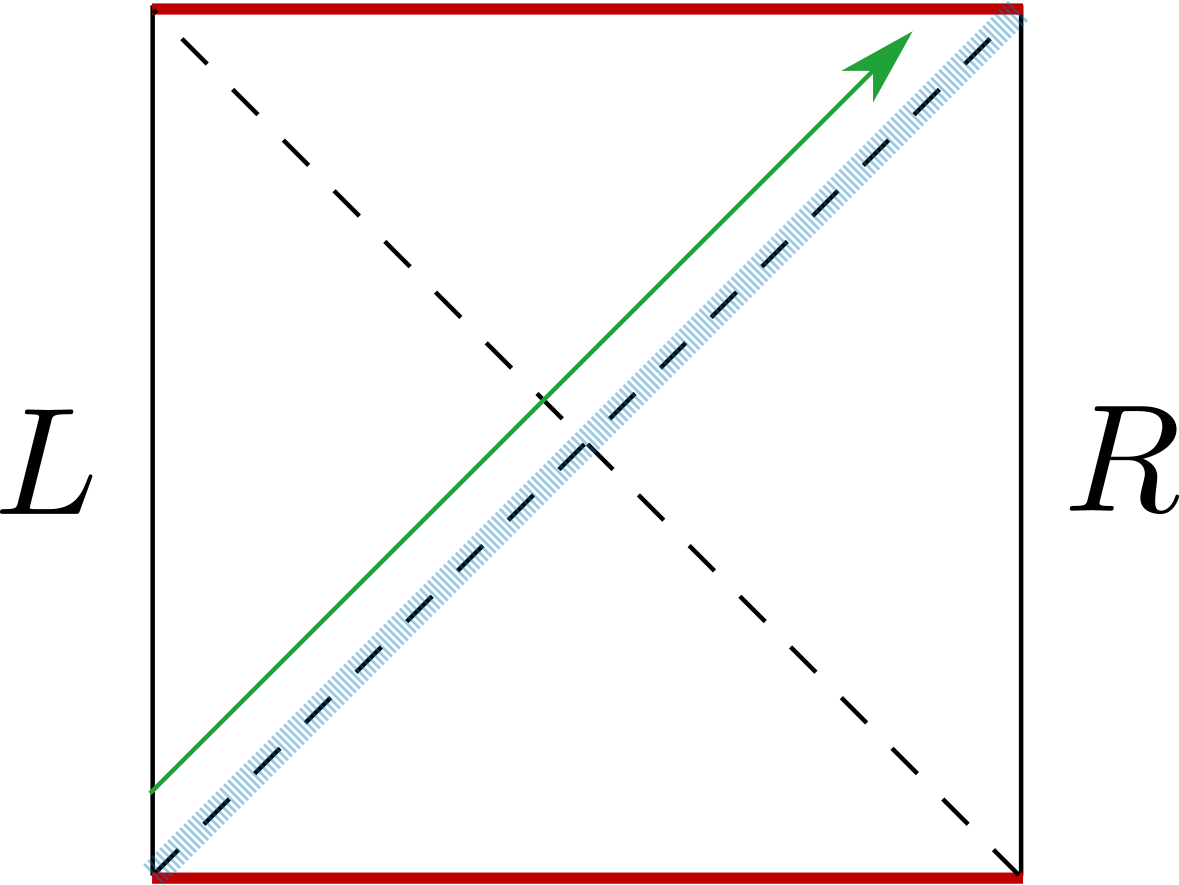}
\caption{Geometry of the AdS wormhole dual to two entangled CFTs living on the boundaries. The ANEC along the horizon (shaded blue) guarantees that a signal sent from the left boundary (green arrow) does not traverse the wormhole and crashes into the singularity.}
\label{fig:wormhole}
\end{figure}

The motivation for the bound discussed above originates from considering evaporating black holes in AdS \cite{Almheiri:2018xdw}. Take the two sided eternal AdS black hole or the wormhole connecting two asymptotically AdS regions. In the context of AdS/CFT, this geometry can be understood as the holographic dual of a pair of decoupled but entangled CFTs prepared in the thermofield double state. We will label these two boundary CFTs as $L$ (left) and $R$ (right), see figure \ref{fig:wormhole}.

The decoupling of the two CFTs should be manifested in the bulk dual as the absence of causal signaling between the two boundaries through the AdS wormhole spacetime. This translates to the wormhole being non-traversable. 
Traversability is precluded by the so-called average null energy condition \cite{Morris:1988tu, Hochberg:1998ii, Visser:2003yf, Visser:2222ft}  on the matter stress energy tensor $T_{ab}$ along the horizon of the black hole
\begin{align}
    \mathrm{ANEC}: \ \ \int T_{a b} k^a k^b d\lambda \ge 0
\end{align}
where $k^a$ is the null tangent vector along the horizon of the black hole and $\lambda$ is an affine parameter along the null ray. The eternal black hole with matter in the Hartle-Hawking vacuum has vanishing stress tensor along its horizons making it only marginally non-traversable. In fact, there is a simple protocol that makes the wormhole traversable by coupling the two boundaries \cite{gao2017traversable}.

We will now consider a setting in which this ANEC places a bound on the energy flux of the boundary system. We will consider the eternal black hole in a 1+1 dimensional setting and allow it to evaporate by imposing absorbing boundary conditions at the boundary. This is a model for starting with two entangled holographic quantum systems dual to the eternal black hole and where one of systems, say the right system, is coupled to an external bath allowing energy to flow between the two.

Consider the Jackiw-Teitelboim (JT) model \cite{Jackiw:1984je, Teitelboim:1983ux, Brown:1988am} coupled to matter given by the action 
\begin{align}
I &= I_{JT}[\phi, g] + I_\mathrm{matter} \\
I_{JT}[\phi, g] &= {1 \over 16 \pi G_N} \int d^2x \ \phi  \left( R+ 2 \right) + {\phi_b \over 8 \pi G_N} \int_{\partial} K
\end{align}
where $\phi$ is the dilaton and $g$ is the two dimensional metric.\footnote{Note we are disregarding a topological term $\phi_0 \int d^2 x R$ in the action which is not important for questions we are interested in regarding dynamics.} This model has been studied recently in \cite{Jensen:2016pah, Maldacena:2016upp, Engelsoy:2016xyb}. Along with this action this theory comes with a pair of boundary conditions on the metric and dilaton
\begin{align}
g_{uu} \sim {1 \over \epsilon^2}, \ \ \phi = \phi_b \sim {\phi_r \over \epsilon}
\end{align}
where $u$ is the time along the boundary and $\epsilon$ is the radial coordinate distance away from the boundary. $u$ is sometimes called the physical boundary time. Integrating over the dilaton along an imaginary contour fixes the metric to be that of AdS$_2$, in which it is convenient to work in Poincare coordinates
\begin{align}
ds^2 = {- dt^2 + dz^2 \over z^2} = {- 4 dx^+ dx^- \over (x^+ - x^-)^2}, \ \ x^\pm = t \pm z.
\end{align} 
The gravitational constraints of this theory imply that the only dynamical gravitational degree of freedom lives on the boundary of the spacetime, and is given by the reparameterization between the Poincare time $t$ and physical boundary time $u$, $t(u)$.

The ADM energy of the spacetime, or the energy as measured on the boundary, is given by
\begin{align}
E = {\phi_r \over G_N} \{ t(u), u \}
\end{align}
The equation of motion of this theory comes from balancing the fluxes of energy of the gravitational sector and the matter.
\begin{align}
 \dot{E}= \dot{t}^2(u) \left[ T_{x^- x^-} - T_{x^+ x^+} \right] \big|_\mathrm{boundary}
\end{align}
where on the right hand of the equation we have the expectation value of the stress tensors evaluated on the boundary of the spacetime.

The eternal black hole is a vacuum solution of this model with
\begin{align}
 T_{x^+ x^+}  =  T_{x^- x^-}  = 0
\end{align}
This fixes the reparameterization, up to an SL2(R) transformation, to be
\begin{align}
t(u) = {\beta \over \pi} \tanh \left[ {\pi \over \beta} u \label{reparam} \right]
\end{align}

Now lets imagine coupling the right boundary to a large external system in the vacuum to allow the black hole to evaporate.  This will only modify the left moving stress $T_{x^- x^-}$. The equation of motion will therefore be
\begin{align}
 \dot{E} = \dot{t}^2(u) T_{x^- x^-}\big|_\mathrm{boundary}
\end{align}
 We want to use this expression to find the stress tensor on the horizon. In general, the relation between the stress tensor near the boundary and the one at the horizon is very complicated in the presence of massive matter or graybody factors. We specialize to the case with matter where these complications are absent, for example by considering conformal matter in the bulk on the background metric $g$, with no coupling to the dilaton. Due to holomorphic factorization, we can write
\begin{align}
    T_{x^- x^-}(x^-) = T_{x^- x^-}(x^-)\big|_\mathrm{boundary} = \dot{t}^{-2}(u) \dot{E}(u)
\end{align}
where we used that $x^- = t$ at the boundary. The first equality follows because there is no dependence on $x^+$ and the stress energy flows on null lines.

We need to plug this into the average null energy along the future horizon on the right exterior. It is important here that we are working to leading order in the gravitational coupling $ \kappa_g \equiv \beta G_N / \phi_r$, so that the reparameterization $t(u)$ is still given by the unperturbed form \eqref{reparam}. Therefore, the horizon along which we want to evaluate the ANEC is still at $x^+ = \beta/\pi$. Using the affine parameterization along the horizon given by
\begin{align}
    x^-(\lambda) &= {\beta \over \pi} - {1 \over \lambda}, \ x^+ = {\beta \over \pi} \\
    k^- &= {d x^- \over d\lambda}, \ k^+ = 0
\end{align}
we have
\begin{align}
    \int T_{a b} k^a k^b d\lambda &= \int T_{x^- x^-} \left( {d x^- \over d \lambda} \right)^2 d\lambda \\
    &= \int_{0}^\infty \dot{E} \ e^{- {2 \pi u  \over \beta}} du
\end{align}
Therefore in this case, the ANEC can be recast as a bound on the integrated energy flux,
\begin{align}
    \mathrm{ANEC} \implies \int_{0}^\infty \dot{E} \ e^{- {2 \pi u  \over \beta}} du \ge 0.
\end{align}
We see that the ANEC translates to a weighted integral of the energy flux at the boundary. This weighting factor is what allows the initial positive energy excitation to overwhelm the subsequent negative energy flux from the black hole losing energy to the external bath. It is interesting that this condition is implied by the general perturbative bound Eq.~\eqref{the_bound}.

\section{Thermalization in SYK}

\subsection{SYK Setup}
\label{sec:syk}

In this section we will study the conventional SYK model~\cite{kitaev,ms}. Let us briefly summarize the relevent results about the conformal limit, Dyson--Schwinger and Kadanoff--Baym equations and also introduce our notations.

SYK is a model of $N$ Majorana fermions $\psi_i$ with the all-to-all interactions and a quench disorder governed by the Hamiltonian:
\beq
H_S = i^{q_S/2}\sum_{1 \leq i_1 < \dots < i_{q_S} \leq N} J_{i_{1} \dots i_{q_S}} \psi_{i_1} \dots \psi_{i_{q_S}}
\eeq
Coefficients $J$ are real and Gauss-random with variance:
\beq
\bra J_{i_{1} \dots i_{q_S}}^2 \ket = \frac{J_S^2 (q_S-1)!}{N^{q_S-1}} \quad \text{(no sum)}
\eeq
Below we will use the symbol $\{i\}$ to denote sums like $1 \leq i_1 < \dots < i_{q_S} \leq N$

Since we are dealing with the quench disorder we have to introduce replicas in the path integral.
However, in the large $N$ limit the interaction between the replicas is suppressed and in the replica-diagonal phase (non-spin glass state) the \emph{Euclidean} effective action reads:
\beq
\label{eq:s_sigma_g}
\frac{S[\Sigma_S,G_S]}{N} = \log \Pf \l \pr_\tau - \Sigma_S \r + \frac{1}{2} \int d\tau_1 d\tau_2 \ \l -\Sigma_S(\tau_1,\tau_2) G_S(\tau_1,\tau_2)  + \frac{J_S^2}{q} G_S(\tau_1,\tau_2)^q \r
\eeq
The auxiliary variables $\Sigma_S, G_S$ have physical meaning: $G_S$ is the Euclidean time-ordered fermion two-point function,
\beq
G_S(\tau_1,\tau_2) = \frac{1}{N} \sum_{i=1}^N \bra T \psi_i(\tau_1) \psi_i(\tau_2) \ket,
\eeq
and $\Sigma_S$ is the fermion self-energy. The large $N$ Euclidean saddle-point equations are identical to the Dyson--Schwinger equations:
\begin{align}
\label{eq:ds}
    \Sigma_S(\tau) = J_S^2 G_S(\tau)^{q_S} \nonumber \\
    (-i \omega - \Sigma_S(\omega)) G_S(\omega) = 1.
\end{align}
For later use, note that the energy is
\begin{equation}
    \label{energy:eucl}
    E_S = \bra  H_S \ket = -\frac{J_S^2}{q_S} \int_0^\beta 
    d \tau \ G_S(\tau)^{q_S}.
\end{equation}

At low temperatures the system develops an approximate conformal symmetry and the Green's function can be found explicitly:
\beq
G_S(\tau) = b \sgn(\tau) \l \frac{\pi}{\beta J_S \sin \l \frac{\pi |\tau|}{\beta} \r} \r^{2/q}
\eeq
The coefficient $b$ is just a numerical constant. It is determined by
\beq
b^{q_S} \pi = \l \oh - \frac{1}{q_S} \r \tan \frac{\pi}{q_S},
\eeq
and for $q_S=4$, $b=1/(4 \pi)^{1/4}$.  

Having reviewed the Euclidean properties, let us turn to the Lorentzian (real time) physics. It is be convenient to work with the Keldysh contour right away, so we will assume that the reader is familiar with this technique. 

One central object is the Wightman (or ``greater'') Green's function:
\beq
G_S^>(t_1,t_2) = G_S(t_1^-,t_2^+) = -i \frac{1}{N} \sum_i \bra \psi_i(t_1^-) \psi_i(t_2^+) \ket
\eeq
Please note that we have $-i$ in our definition. Because of the Majorana commutation relations, the greater Green's function reduces to $-i/2$ at coincident points:
\beq
\label{eq:g0}
G_S^>(t_1,t_1) = -\frac{i}{2}.
\eeq
The ``lesser'' function $G^<$ for Majorana fermions is directly related to $G^>$:
\beq
G_S^<(t_1,t_2) = G_S(t_1^+,t_2^-) = - G_S^>(t_2,t_1)
\eeq
Two final pieces are the retarded and advanced functions:
\begin{align}
    G_S^R(t_1,t_2) = \theta(t_1-t_2) \l  G_S^>(t_1,t_2) - G_S^<(t_1,t_2)\r \nonumber \\
    G_S^A(t_1,t_2) = \theta(t_2-t_1) \l G_S^<(t_1,t_2) - G_S^>(t_1,t_2) \r
\end{align}

Now we can finally write down the Lorentzian form of the Dyson--Schwinger equations (\ref{eq:ds}). The analytic continuation of the time-ordered Euclidean Green's function from imaginary time to real time yields the
Wightman function, and because the self-energy equation is naturally formulated in real time, one gets
\begin{equation}
\label{eq:kb1}
    \Sigma_S^>(t_1,t_2) = - i^{q_S} J_S^2 \l G_S^>(t_1,t_2) \r^{q_S}.
\end{equation}
However, the continuation in frequency space (from the upper-half plane) yields the retarded function, therefore the second equation in (\ref{eq:ds}) transforms into:
\begin{equation}
    G_S^R(\omega) ( \omega - \Sigma^R(\omega)) = 1
\end{equation}
We need a relation between $G^R$ and $G^>$ in order to close the system of equations. If the system is in a thermal state, this relation is provided by the fluctuation-dissipation theorem (FDT):
\beq
G_S^>(\om) = 2 i \Im G_S^R(\om) \frac{1}{e^{-\beta \omega}+1}.
\eeq
This system of equations can be solved by an iterative procedure to obtain the real time correlation functions~\cite{ms,quench1}.

There is another away we can treat the second DS equation in (\ref{eq:ds}). We can rewrite it in the time-domain using the
convolution:
\beq
\pr_\tau G_S(\tau) = \delta(\tau) + \int_0^\beta d\tau' \ \Sigma_S(\tau-\tau') G_S(\tau').
\eeq
Upon the analytic continuation this yields the so-called Kadanoff--Baym equations\footnote{The integral on the
right hand side is simply the convolution along the Keldysh contour of $\Sigma_S(t_1^+,\cdot) * G_S(\cdot,t_2^-)$.
The precise result for the integral is known as Langreth rule in condensed matter literature \cite{stefanucci2013}.}:
\begin{align}
\label{eq:kb2}
    i \pr_{t_1} G_S^>(t_1,t_2) = \int_{-\infty}^{+\infty} dt' \ \l \Sigma^R(t_1,t') G_S^>(t',t_2) + \Sigma^>(t_1,t') G^A(t',t_2) \r \nonumber \\
    -i \pr_{t_2} G^>(t_1,t_2) = \int_{-\infty}^{+\infty} dt' \ \l G^R(t_1,t') \Sigma^>(t',t_2) + G^>(t_1,t') \Sigma^A(t',t_2) \r.
\end{align}
Note that these equations are causal due to the retarded and advanced propagators in the integrand. A more straightforward way to obtain these equations is to write down the large $N$ effective action (\ref{eq:s_sigma_g}) on the Keldysh contour and find the classical equations of motion.\footnote{Strictly speaking, for a thermal initial state the right hand side contains the integral over the imaginary axis running from $-i \beta$ to $0$. However we can imagine that all non-equilibrium processes happen at large positive Lorentzian times so this piece is
essentially zero if correlators decay with time. This is the reason why the integration over $t'$ starts at $-\infty$.
} 

These equations can be used in non-equilibrium situations. They also have a very generic form that simply encodes the relation between the Green's function and the self-energy. So, the actual non-trivial piece of information is the relation (\ref{eq:kb1}). When we couple the system to a bath the integral equations (\ref{eq:kb2}) will stay 
exactly the same, whereas the answer for the self-energy (\ref{eq:kb1}) will change. Appendix~\ref{kb:numerics}
describes our approach to solving the KB equations (\ref{eq:kb1}).

To conclude this subsection, let us write down the expression for the energy:
\beq
\label{energy:lorentz}
E_S(t) = \bra H_S(t) \ket = -i^{q_S+1} \frac{J_S^2}{q_S} \int^t_{-\infty} 
dt' \ \l G_S(t,t')^{q_S} - G_S(t',t)^{q_S} \r
\eeq
Using the equations of motion (\ref{eq:kb2}), it follows that 
\beq
\label{eq:e_dg}
E_S(t) = \frac{1}{q_S} \pr_t G_S(t,t).
\eeq
This is not a general expression; it holds because the SYK Hamiltonian only contains terms with $q_S$ identical fermions.

\subsection{Coupling to a bath}
\label{sec:coupling_to_bath}

Suppose that one system fermion $\psi$ couples to an external bath operator $\Oc_B$,
\beq
S_{\rm int} = i V \int_\Cc du \psi \Oc_B.
\eeq
If the interaction is weak enough, we can use the 1-loop approximation
to the interaction term:
\begin{align}
\label{int:1loop}
S_{\rm int} \rightarrow \frac{V^2}{2} \int_\Cc du_1 du_2 \  X_B(u_1,u_2), G_S(u_1,u_2)  
\end{align}
where the function $X_B$ is simply the two-point function of the bath operator 
$X_B(t_1,t_2)=\bra \Oc_B(t_1) \Oc_B(t_2) \ket_B$. Moreover, if the bath is large, we can neglect the back reaction on the bath and take $X_B$ to be fixed. 
This logic can be made precise if we couple a large-$N$ SYK to another large-$M$ SYK with $M \gg N$. Consider a general interaction of the form\footnote{A similar
interaction was independently studied in \cite{Chen2017,zhang2019evaporation}}
\beq
\label{v:def}
V^{j_1, \dots, j_{f_B}}_{i_1, \dots, i_{f_S}} \psi_{i_1} \dots \psi_{i_{f_S}} \chi_{i_1} \dots \chi_{i_{f_B}},
\eeq
where $\chi_j,\ j=1,\dots,M$ are the $M$ Majorana fermions of the bath and $V^{j_1, \dots, j_{f_B}}_{i_1, \dots, i_{f_S}}$ 
is a random Gaussian variable with variance
\beq
\bra  \l V^{j_1, \dots, j_{f_B}}_{i_1, \dots, i_{f_S}} \r^2 \ket = \frac{V^2 (f_S-1)! f_B!}{N^{f_S-1} M^{f_B}} \quad \text{(no sum)}.
\eeq
Note that this expression allows for a quite general coupling between $f_S$ system fermions and $f_B$
bath fermions. Based on it, we can derive an effective action similar to (\ref{eq:s_sigma_g}). 
The Euclidean action has the form:
\begin{align}
S &= \sum_{i=1}^N \psi_i \pr_\tau \psi_i - i^{q_S/2} \sum_{\{i\}} J_S^{\{i\}} \psi_{i_1} \dots \psi_{i_{q_S}} \nonumber \\
& + \sum_{j=1}^M \chi_j \pr_\tau \chi_j  -  i^{q_B/2} \sum_{\{j\}} J_B^{\{j\}} \chi_{j_1} \dots \chi_{j_{q_B}} \nonumber \\
& - i^{\gamma}\sum_{\{i\},\{j\}}V^{\{j\}}_{\{i\}} \psi_{i_1} \dots \psi_{i_{f_S}} \chi_{i_1} \dots \chi_{i_{f_S}},
\end{align}
where 
\begin{align}
    \gamma = f_S f_B + f_S(f_S-1)/2 + f_B(f_B-1)/2.
\end{align}
The powers of $i$ are needed to make the action real.

It is convenient to introduce the Green's functions by adding Lagrange multipliers $\Sigma_S, \Sigma_B$ (which we integrate over the imaginary axis):
\begin{align}
\Delta S & = \frac{1}{2}\Sigma_S \l \sum_{i} \psi_i(\tau_1) \psi_i(\tau_2) - N  G_S(\tau_1,\tau_2)   \r \nonumber \\ 
& + \frac{1}{2}\Sigma_B \l \sum_{j} \chi_i(\tau_1) \chi_i(\tau_2) -M  G_B(\tau_1,\tau_2)   \r
\end{align}

If we assume no replica symmetry breaking, we can treat $J_S, J_B, V$ as conventional integration variables in the path integral. After integrating them out, we can replace fermionic bilinears with $G_{S/B}$. Then the action becomes quadratic in $\psi$ and $\chi$ and they can be integrated out as well. The result is the following effective action:
\begin{align}
S &= N \log \Pf \l  \pr_\tau - \Sigma_S \r + M \log \Pf \l  \pr_\tau - \Sigma_B \r \nonumber \\
&+ \frac{1}{2} \int d\tau_1 d\tau_2  \ \l - N \Sigma_S G_S -  M \Sigma_B G_B +
\frac{N}{q_S} G_S^{q_S} + \frac{M}{q_B} G_B^{q_B} \r  \nonumber  \\
 &+ \frac{V^2 N}{2 f_S} \int d\tau_1 d\tau_2 \ G_S^{f_S} G_B^{f_B}.
\end{align}
Hence, the equations of motion for the bath variables $G_B,\Sigma_B$ will be corrected by
a term of order $N/M$ which is suppressed for $M \gg N$. However, there is a non-vanishing correction
the the system's self-energy:
\begin{align}
    \Sigma_S = J_S^2 G_S^{q_S-1} + \Delta \Sigma_S \nonumber \\
    \Delta \Sigma_S = V^2 G_S^{f_S-1} G_B^{f_B}.
\end{align}

The same computation can be performed in Lorentzian time with the following result for $\Delta \Sigma_S^>$:
\beq
\label{s_extra}
\Delta \Sigma^>_S(t_1,t_2) = - i^{f_S+f_B} \l G^>_S(t_1,t_2) \r^{f_S-1} \l G^>_B(t_1,t_2) \r^{f_B}.
\eeq

Computations in SYK simplify a lot when we take the large $q$ limit. Now we have additional parameters $f_{S/B}$ which we can take to infinity along with $q_{S/B}$. For example, consider large $q_B$. Then the Euclidean Green's function has the following expansion:
\beq
G_B(\tau) = \frac{1}{2} \sgn(\tau) \l  1  + \frac{g_B}{q_B} \r
\eeq
By taking $q_B \rightarrow +\infty, f_B \rightarrow +\infty$ we get the following term in the interaction:
\beq
G_B^{f_B} = \const \ \exp \l \frac{f_B}{q} g_B \r,
\eeq
where $f_B/q$ can be any rational number. Recall that at zero temperature and large $q_B$ one has \cite{ms}:
\beq
e^{g_B} = \frac{1}{(J_B |\tau| +1)^2}.
\eeq
This provides an example when we know the bath Green's function explicitly for all times. 

\subsection{Equilibrium}

Let us first study the equilibrium Dyson--Schwinger equation in presence of a bath. It happens that they can be solved in the IR regime. With the above setup, the Euclidean self-energy reads:
\beq
\Sigma_S = J_S^2 G_S^{q_S-1}  +  V^2  G_S^{f_S-1} G_B^{f_B}.
\eeq
In equilibrium, the system and bath will have the same temperature. Thus, we make an ansatz for $G_S$ which is an SYK Green function with certain effective parameters $\tilde{q_S}, \tilde{J_S}$.

Suppose first that we try to retain the same $q_S$, so $\tilde{q_S}=q_S$. Remember that the SYK Green functions decay for large Lorentzian times as $G_S \sim 1/\sinh(\pi t/\beta)^{2/q_S}$. Thus, there are three possible situations:
\begin{itemize}
\item The system term ($G_S^{q_S-1}$) dominates in the IR. Then the 
interaction with the bath is irrelevant and in the IR we recover the decoupled system physics.
\item The bath term ($G_S^{f_S-1} G_B^{f_B}$) dominates. This means $G_S$ with $\tilde{q_S}=q_S, \tilde{J_S}=J_S$ is not a solution. The interaction is relevant and the system now has an effective $\tilde{q_S}<q_S$. The solution can be found by assuming that for a given $\tilde{q_S}$ the second term is dominant.
\item Both terms are of the same order, so the interaction is marginal. This means that $\tilde{q_S}=q_S$, but $J$ can be renormalized.
\end{itemize}

Let us study the particular example of a marginal deformation of $q_S=4$, where the bath is also a $q_B=4$ SYK. Take $f_S+f_B=4$. In Euclidean time the full DS equation reads:
\beq
\Sigma = J_S^2 G_S^3 + V^2 G_S^{f_S-1} G_b^{4-f_S},
\eeq
and in the low energy limit,
\beq
\Sigma * G_S = \delta(t_1-t_2).
\eeq 
Recall also that $G_B$ satisfies the following equation in the IR:
\beq
J_B^2 G_B^3 * G_B = \delta(t_1-t_2)
\eeq

The ansatz is then $G_S = \frac{\sqrt{J_B}}{\sqrt{\tilde{J_S}}} G_B$. Remembering that in the IR the only dependence on the coupling is $G_B \sim 1/\sqrt{J_B}$, this ansatz can be understood as a renormalization of the quartic SYK coupling. 

From this ansatz it follows that $\tilde{J}$ is determined by,
\beq
\label{ren:fin_q}
\frac{J_S^2}{\tilde{J_S}^2} + V^2 \frac{J_B^{f_S/2-2}}{\tilde{J_S}^{f_S/2}}=1.
\eeq
This corresponds to an increase in the effective coupling $\tilde{J}_S$ relative to $J_S$. We have also confirmed this equation in our numerical results.

\subsection{Energy flux}

We now derive an equation for the rate of change of the system energy within the non-equilibrium formalism. Suppose we have a generic quantum system with a Hamiltonian $H_S$, which we couple to a bath with
Hamiltonian $H_B$ and the interaction term is $V \Oc_B \Oc_S$, which we turn on at time $t=0$. The total Hamiltonian
is
\beq
H=H_S + H_B + V \theta(t) \Oc_S \Oc_B
\eeq
The time derivative of the system's energy is not zero for $t>0$:
\beq
E_S'=\frac{d}{dt} H_S = -i [H,H_S] =  V \pr_t \Oc_S \Oc_B,
\eeq
where the $\pr_t \Oc_S$ indicates the time derivative of this operator with respect to unperturbed equations of motion for the system.
If the bath is large, we completely ignore the back-reaction on the bath. Moreover, if $V$ is small, we can find the
right hand side in perturbation theory in $V$:
\beq
\label{flux:general}
E_S' = -i (-1)^F \int_\Cc dt' \pr_t \bra V^2  \Oc_S(t) \Oc_S(t') \ket_S  \bra \Oc_B(t) \Oc_B(t') \ket_B,
\eeq
where the integral over $t'$ goes along the Keldysh contour from $0$ to $t$ and the correlators are taken in the unperturbed systems. $F$ is the fermion number of the operator $\Oc_S$.
For the SYK model with a random interaction (\ref{v:def})
this equation leads to 
\beq
\label{flux:syk}
E_S' = i V^2 \int_{-t}^t du G^>_B(u-i \ep)^{f_B} 
\pr_u G^>_S(u- i \ep)^{f_S}.
\eeq
In this specific case, this equation can also be derived directly from the Kadanoff--Baym equations (Appendix \ref{flux_kb}) or from Schwarzian (Section \ref{late:sch}).

\subsection{Very early time}

At very early times, $t \ll 1/J_S, 1/J_B$, we can assume that 
$G_B$ and $\pr_u G_S$ are just constants. Then we can use the relation
(\ref{eq:g0}) for $G_B$ and Eq. (\ref{eq:e_dg}) to connect the derivative with system's energy. Collecting the factors of $i$, we obtain:
\beq
E_S'=-t  V^2 E_S(0) \frac{1}{2^{f_S+f_B-2}}.
\eeq
Since for SYK $E<0$ in thermal equilibrium, we see that the energy initially increases. This is an illustration of the general statement we discussed in Section~\ref{sec:bound}. In the case of SYK, the initial energy growth rate is proportional to the initial energy. This result is very general for SYK and an arbitrary bath of Majorana fermions. This is valid for SYK at any coupling $J_S \beta$. 

\subsection{Early time}

At early times, the state of the
system has not changed much, so we can use the initial $G_S$ in Eq. (\ref{flux:syk}). Put another way, Eq. (\ref{flux:syk}) is already the leading term in $V^2$, so $V$-corrections to $G_S$ are smaller. We can trust this approximation as long as change in $\beta$ is of order $V$. Below we will argue that we can use the conformal approximation for $G_S$, so we must restrict ourselves to times $t \gtrsim 1/J_S$.

Also, from now on we assume that the system is $q_S=4$ SYK and there is only one system fermion in the interaction, $f_S=1$. By going to the conformal limit, we will arrange the situation so that an analytical calculation of various parts of the energy curve is possible. This is also the limit of interest for the black hole evaporation problem. Since the bath temperature is set to zero, we now denote $\beta_S$ by just $\beta$.

For finite $q$ SYK we know the Green's function analytically only in the conformal regime, when $u \gg 1/J_B$. If we try to use the conformal answer for $G_B$ we will encounter a divergence at $u=0$ for
$f_B \ge 2$, since at short times $G_B \sim 1/\sqrt{u}$. The physically interesting cases of marginal and irrelevant deformations correspond to $f_B \ge 3$ so we need to find another approximation to $G_B$. One way around this is to couple the system to large-$q$ SYK. For simplicity we will study the case of zero temperature bath. As we have shown in Section \ref{sec:coupling_to_bath}, we can adjust the number of bath fermions in the interaction such that 
\beq
G_B^{f_B} = \l \frac{i}{2}  \frac{1}{(i J_B u + 1)^{1/2}} \r^p
\eeq
for any rational number $p$. Then $p=3$ corresponds to a marginal deformation, whereas $p>3$ produces an irrelevant interaction.

With this setup, the relevant integrals converge. However, the question of whether we can use the conformal approximation for $G_S$ is
still open. One obvious constraint is that the system is at strong coupling, so we require
\beq
J_S \beta \gg 1
\eeq
From the integral in Eq. (\ref{flux:syk}) it follows that the bath probes the system Green's function at times of order $1/J_B$. In order to use the 
conformal approximation for $G_S$, this time should be large compared to $1/J_S$. So we should restrict ourselves to 
\beq
J_S \gg J_B.
\eeq

In the conformal approximation,
\beq
G_S = b \l i\frac{\pi}{\beta J_S \sinh(\pi u/\beta)} \r^{1/2},
\eeq
so $\pr_u G_S$ contains a $1/u^{3/2}$ term at small $u$ which generates divergences in integrals. However, we can integrate by parts to give,
\beq
\label{eq:part_int}
E_S' = -i V^2 \l G_B^{f_B}(t) G_S(t) - G_B(-t)^{f_B} G_S(-t) - \int_{-t}^{t} du \ \pr_u G_B(u)^{f_B} G_S(u-i \ep) \r.
\eeq
Using the fact that for Majorana fermions the Green's function obeys
\beq
G_{S,B}(-t)=-G_{S,B}^*(t),
\eeq
we can rewrite the flux as
\beq
\label{eq:part_int_im}
E_S' =  2 V^2 \ \Im  \l G_B^{f_B}(t) G_S(t) - \int_{0}^{t} du \ \pr_u G_B(u)^{f_B} G_S(u-i \ep) \r. 
\eeq

\begin{figure}[ht!]
\centering
\includegraphics[scale=1.0]{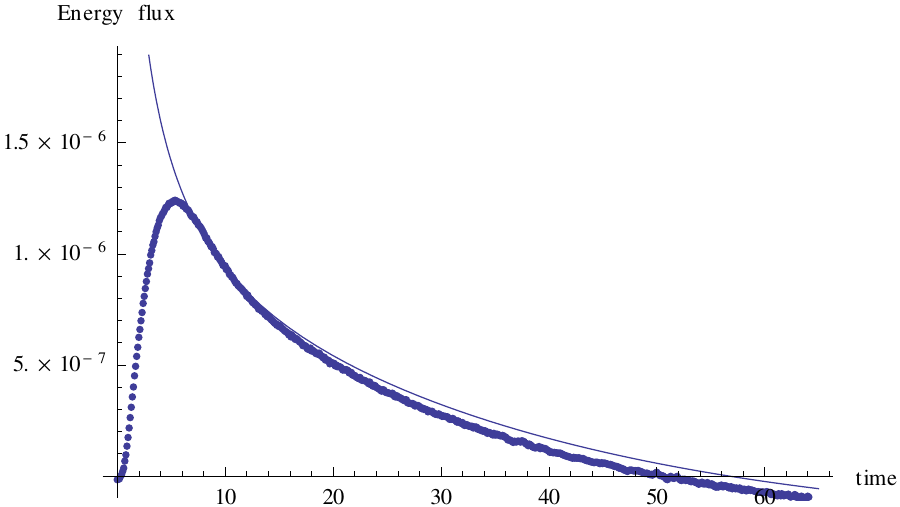}
\caption{Early time energy flux as function of time. The solid curve is the analytic result (\ref{eq:part_int}), and the dots show the direct numerical integration. The parameters used are 
$V^2= 2.5 \times 10^{-5}, J_S=0.5, J_B=0.005, \beta_{\rm init}=50, dt=0.1$. The conformal approximation is responsible for the disagreement at early time while the slight change in temperature is responsible for the disagreement at late time.}
\label{fig:sch_early}
\end{figure}

A comparison between this result and the exact numerical integration for the marginal case $p=3$ is presented in Figure \ref{fig:sch_early}. Notice that the two curves do not quite match
at very early times. Because of the form of the conformal propagator, the flux behaves as $1/\sqrt{t}$, which
is not physical. Had we taken the exact system two-point function we would have reproduced the numerical answer perfectly even at very early times. Slight deviations occur later because the system's temperature is finally changing. These discrepancies decreases with decreasing the system-bath coupling.

We have studied the analytic expression for the marginal case $p=3$ in two limits, $J_B \beta \ll 1$ and $ J_B \beta \gg 1$, in Appendix \ref{app:peak}. The parameter $\beta J_B$ tells us how ``fast'' the bath degrees
of freedom are compared to the thermal scale of the system. For a ``slow bath'' with $J_B \beta \ll 1$, the peak occurs at times logarithmically bigger than $\beta$:
\beq
t_{\rm peak} \sim \beta \log \l \frac{1}{J_B \beta} \r, J_B \beta \ll 1.
\eeq
In the opposite limit of a ``fast bath'' with $J_B \beta \ll 1$, we find that the peak time is much less than $\beta$:
\beq
t_{\rm peak} \sim \frac{\beta}{\l \beta J_B \r^{1/3}} , J_B \beta \gg 1.
\eeq

\subsection{Intermediate time}
\label{late:sch}

At finite temperature the Green's functions in Eq. (\ref{flux:syk}) decay exponentially with time. Assuming that the bath is at a lower temperature than the system, the integral saturates at times $t \gtrsim \beta$. After this the energy flow comes to a steady state, meaning that it is not sensitive to when exactly the interaction was switched on. 

We can patch this regime with the previous discussion if the coupling is small enough. Namely the change in temperature over thermal time scale is much less than temperature:
\beq
\beta \partial_u \beta \ll \beta \rightarrow \partial_u \beta \ll 1.
\eeq
But this requirement is equivalent to saying that at each point in time the system is in quasi-equilibrium and has a definite temperature.

We expect that in this regime the system's 
dynamics can be described by the Schwarzian.
In this approximation, the system's Lagrangian is equal to the Schwarzian derivative:
\beq
S_\text{kin}=- \frac{\al_S}{\Jc_S} \int_\Cc du \ \Sch(t,u)=\frac{\al_S \sqrt{2}}{2 J_S} 
\int_\Cc du \ \l  \frac{t'''}{t'}  - \frac{3}{2} \l \frac{t''}{t'} \r^2 \r.
\eeq
Since for $q=4$, $\Jc_S=J_S/\sqrt{2}$ and the coefficient in front of the Schwarzian is $\al_S=0.007$. 
The system's energy is given by,
\beq
\label{energy:sch}
E=E_0+ \frac{\al_S \sqrt{2}}{J_S}  \Sch(t[u],u) = E_0 + \frac{2 \pi^2 \sqrt{2} \al_S}{J_S \beta^2}, 
\eeq
where $E_0$ is the ground state energy.

The interaction with the bath comes from reparametrizations
of $G_S$ in the action term (\ref{int:1loop}):
\begin{align}
\label{sch:int}
S_\text{pot} = i \frac{V^2}{2} \int_\Cc du_1 du_2  X_B(u_1,u_2) G_S (u_1,u_2)  = \\ \nonumber 
=  i \sqrt{i} b \frac{V^2}{2 J_S^{1/2}  } \int_\Cc du_1 du_2  
 \l \frac{ t'[u_1] t'[u_2]}{ (t[u_1]-t[u_2])^2  } \r^{1/4} X_B(u_1-u_2).
\end{align}
With this normalization of $V^2$, one has the following extra term in the Dyson--Schwinger equation (compare with Eq. (\ref{s_extra})):
\beq
\Delta \Sigma^>_S = - V^2 X^>_B.
\eeq

Note that the above action is written on the Keldysh contour, so we have two functions $t_+[u],t_-[u]$. The semiclassical equations of motion
are obtained by varying with respect to $t_q = t_+ - t_-$ and putting $t_+=t_-$ \cite{kamenev_book}. This way the equations of motion are causal.

During the approach to equilibrium we expect that the
solution has the form
\beq
t[u]= \tanh \l \frac{\pi u}{\beta[u]} \r
\label{beta_u},
\eeq
where $\beta[u]$ is a slowly varying function of $u$. As discussed, the difference between the times $u_1,u_2$ in 
$G_S(u_1,u_2)$ should be less than the
characteristic scale at which $\beta$ changes: $\beta/\beta' \gg |u_1-u_2|$.

We go through the derivation of the equations of motion in Appendix \ref{sch:eom}. The result is:
\begin{align}
\label{eq:betap}
\frac{4 \pi^2 \sqrt{2} \al_S}{J_S \beta^3} \beta' =     
\frac{i \sqrt{i} b V^2 \pi^{3/2}}{2  (J_S \beta)^{1/2}}  \int_{-\infty}^{+\infty} du \ X_B(\beta(u - i \ep))
\frac{\cosh \pi  \l u - i \ep \r}{\sinh^{3/2} \pi  \l u  - i \ep \r}.
\end{align}
This result coincides with the general answer (\ref{flux:syk}) when the
system Green's function is approximated by the conformal expression and the energy of the system is given by Schwarzian
result (\ref{energy:sch}).

There is one subtlety here.\footnote{We are grateful to Juan Maldacena for a discussion on this point.} In the Schwarzian approximation the energy above the vacuum is proportional to $-\Sch(t[u],u)$. For a thermal state this is equal to $2 \pi^2/\beta^2$. Correspondingly we expect that the energy flux is proportional to $-4 \pi^2 \beta'/\beta^3$. This is how (\ref{eq:betap}) was obtained (see Eq. (\ref{lag_kin}) in Appendix \ref{sch:eom}). However, if we formally evaluate the Schwarzian on the configuration (\ref{beta_u}) we will get an extra term,
\beq
\label{extra_betap}
-\Sch(t[u],u) = \frac{2 \pi^2}{\beta^2} - \frac{4 \pi^2 u \beta'}{\beta^3}.
\eeq
And after differentiating with respect to time $u$, we get an expression $-8 \pi^2 \beta'/\beta^3$, which is twice as big as it should be.

The resolution of this problem is that the expression (\ref{beta_u}) is not an actual solution if $\beta[u]$ is not constant. The argument of $\tanh$ should include an additional term proportional to $\beta'$ in order to cancel the extra derivative term in Eq. (\ref{extra_betap}). The true solution is easily found,
\beq
t[u] = \tanh \l \frac{\pi u}{\beta[u]} + \frac{\pi u^2 \beta[u]'}{2 \beta[u]^2} \r.
\eeq

Now we specialize again to the case of a $q_B=4$ bath and study both marginal and irrelevant interactions. We also assume that the bath is at strong coupling. And by this we mean that it is strongly coupled by itself,
\beq
J_B \beta_{\rm bath} \gg 1,
\eeq
and it is strongly coupled on the thermal time scale of the system,
\beq
J_B \beta \gg 1.
\eeq
Otherwise, the $\epsilon$-prescriptions in integrals should be replaced by the actual UV cut-off $\sim 1/(\beta J_{S/B})$.

In the subsequent sections we are going to compare Schwarzian 
results with numerical computations. Our timestep will be $dt=0.1$ and $J_S=0.5$, so all the numerical answers should
come with $\sim J dt = 0.05 = 5\%$ uncertainty. Later when we check the bound we will estimate the uncertainties more carefully. 

\subsubsection{Marginal deformation: bath at zero temperature}

For a marginal deformation with $f_B=3$ and when the bath is at zero temperature, the function $X_B$ is given by
\beq
X_B =  G_B^3 =  i^{3/2} \l  \frac{b}{ \sqrt{J_B t}}\r^3,
\eeq
and the integral over $x$ evaluates to,
\beq
\label{int1}
\int_{-\infty}^{+\infty} dx \ \frac{1}{ (x - i \ep)^{3/2}} 
\frac{\cosh \pi  \l x - i \ep \r}{\sinh^{3/2} \pi  \l x  - i \ep \r} =  i \frac{\pi^{3/2}}{4}.
\eeq

\begin{figure}[!ht]
\centering
\includegraphics[scale=0.5]{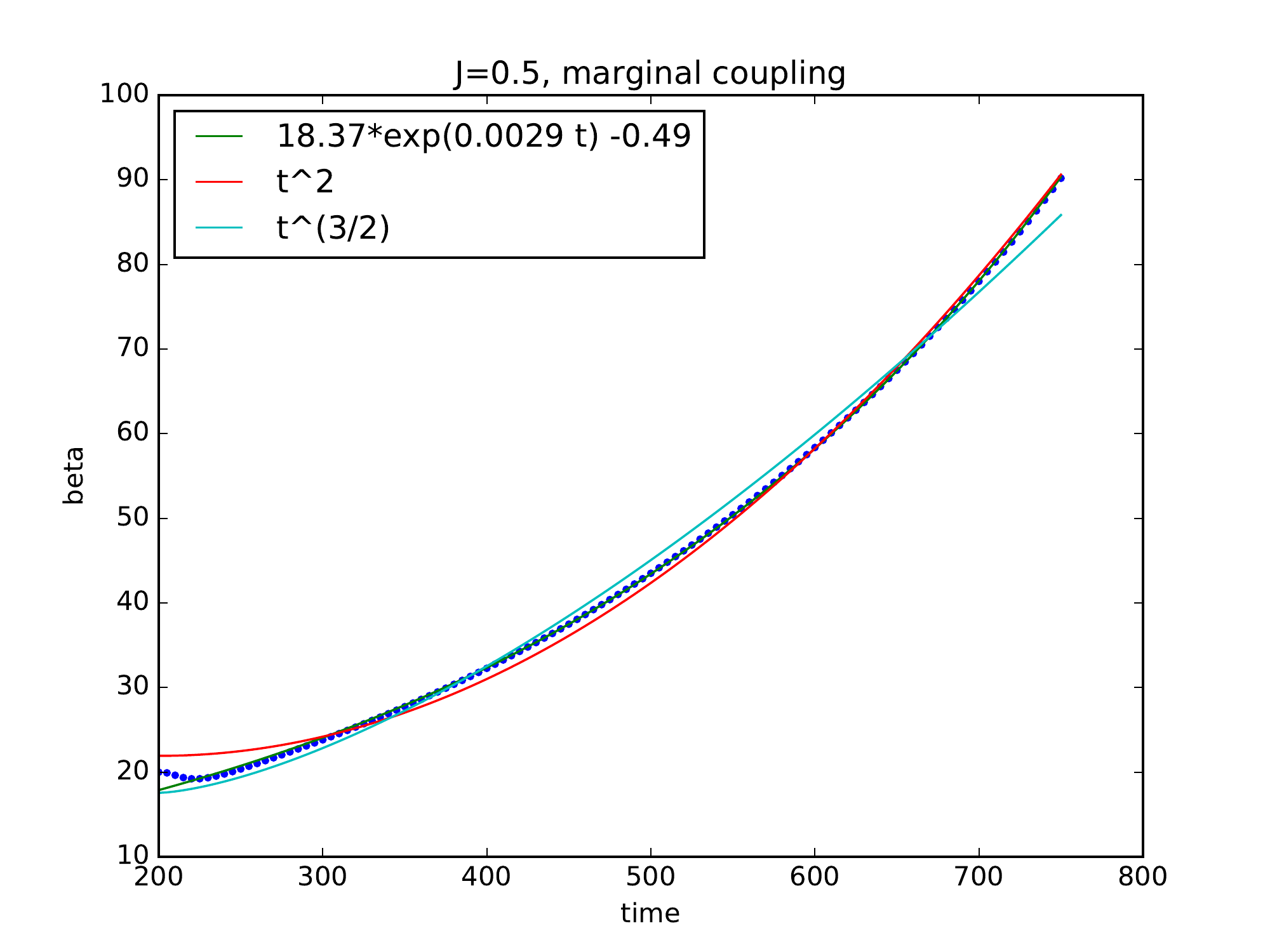}
\caption{$\beta$ as a function of time for a marginal coupling. Blue circles are data points, and the green curve is an exponential fit. The fit almost coincides with data points. Fits for other powers are shown for comparison.}
\label{fig:exp}
\end{figure}

The time-dependence of the temperature is therefore
\beq
\label{betap:early}
\beta'=  \beta  \frac{\pi V^2 b^4 \sqrt{J_S}}{32 \sqrt{2} \al_S J_B^{3/2}}, \quad \text{marginal}.
\eeq
For $J_S=J_B$ we have verified this numerically as shown in Figure \ref{fig:exp}. The above equation yields $\beta' = 0.0028 \beta$ for $V^2=0.002, J=0.5$, whereas the best fit from numerics is 
$\beta'= 0.0029 \beta$.

\subsubsection{Marginal deformation: bath at finite temperature}

\begin{figure}[!ht]
\centering
\includegraphics[scale=0.5]{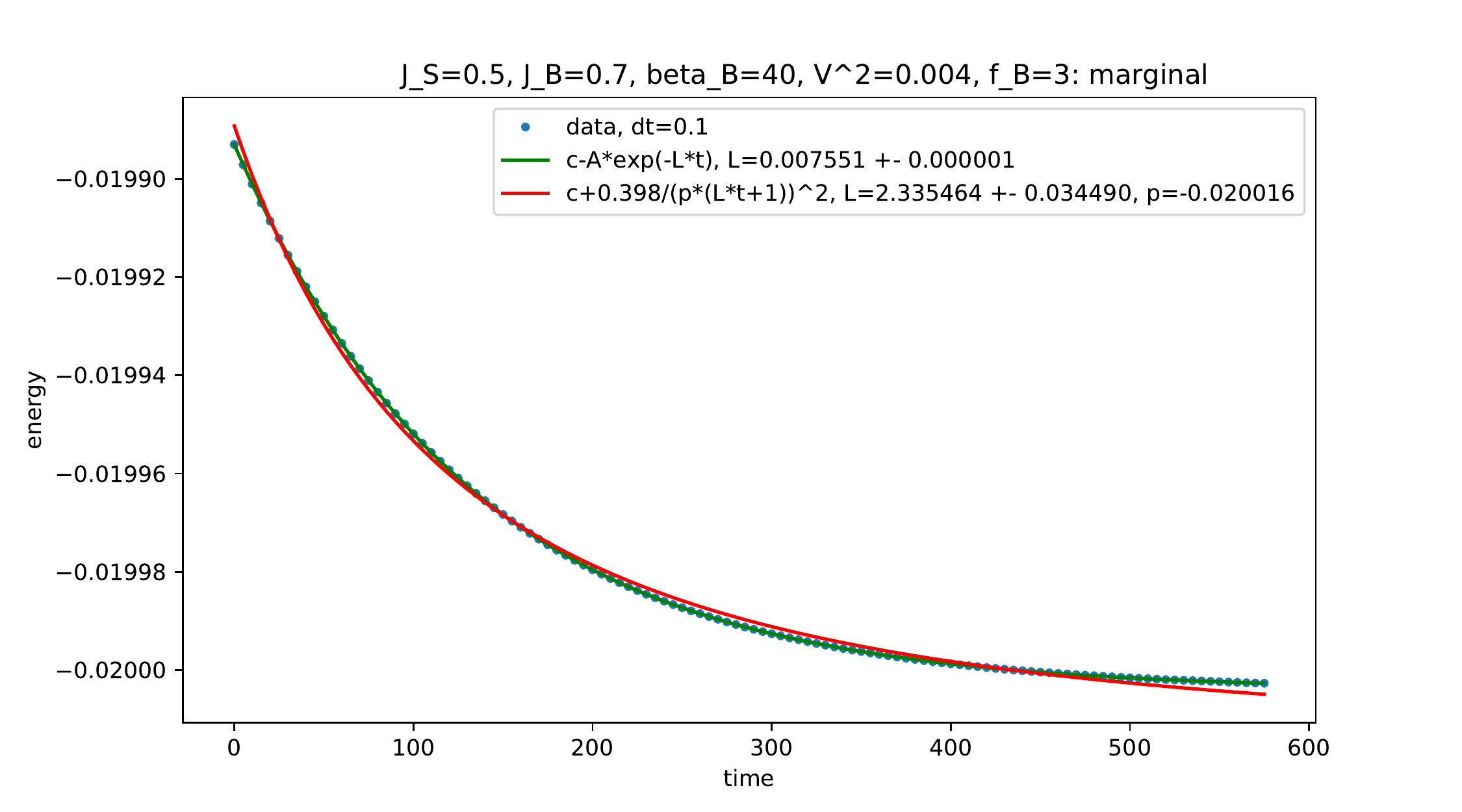}
\caption{Energy as a function of time for $\beta_{\rm init}=35$ and $\beta_B=40$. Only the late time behavior is shown. The green line is an exponential fit. For comparison we
also included fits with $E \sim 1/t^2$. The analytical answer for the rate is $0.0075$ from Eq. (\ref{fin_p3}).}
\label{beta50_p3}
\end{figure}

In this case, we take the conformal SYK answer for the bath Green's function:
\beq
X_B = G_B^3 = i^{3/2} \l  \frac{b \sqrt{\pi}}{ \sqrt{\beta_B J_B \sinh \l \frac{\pi u}{\beta_B} \r}}\r^3.
\eeq
If $\beta_B \sim \beta$, then we can expand the integral in powers of $\beta_B - \beta$:
\begin{align}
\label{int2}
\int_{-\infty}^{+\infty} dx \ \frac{\pi^{3/2} \beta^{3/2}}{ \beta_B^{3/2} \sinh^{3/2} \l \pi (x-i \ep) \frac{\beta}{\beta_B} \r} 
\frac{\cosh \pi  \l x - i \ep \r}{\sinh^{3/2} \pi  \l x  - i \ep \r} = \\ \nonumber
= -\frac{3}{2} \pi^{5/2} \l \frac{\beta}{\beta_B} -1  \r \int_{-\infty}^{+\infty} dx \ \frac{ (x- i \ep) \cosh^2 \pi \l x - i \ep \r }{  \sinh^4 ( \pi (x-i \ep)) }  = \\ \nonumber
= - i \frac{\pi^{3/2}}{2}  \frac{\beta-\beta_B}{\beta_B}.
\end{align}
The approach to the bath temperature is exponential.
Explicitly, we have
\beq
\beta' =(\beta_B-\beta) \frac{\pi b^4}{32 \sqrt{2} \al_S} \frac{V^2 \sqrt{J_S}}{J_B^{3/2}}, \quad \text{marginal}.
\label{fin_p3}
\eeq
Again, this matches perfectly with the numerics as shown in Figure \ref{beta50_p3}.

\subsubsection{Irrelevant deformation: bath at zero temperature}
\begin{figure}[!ht]
\centering
\includegraphics[scale=0.5]{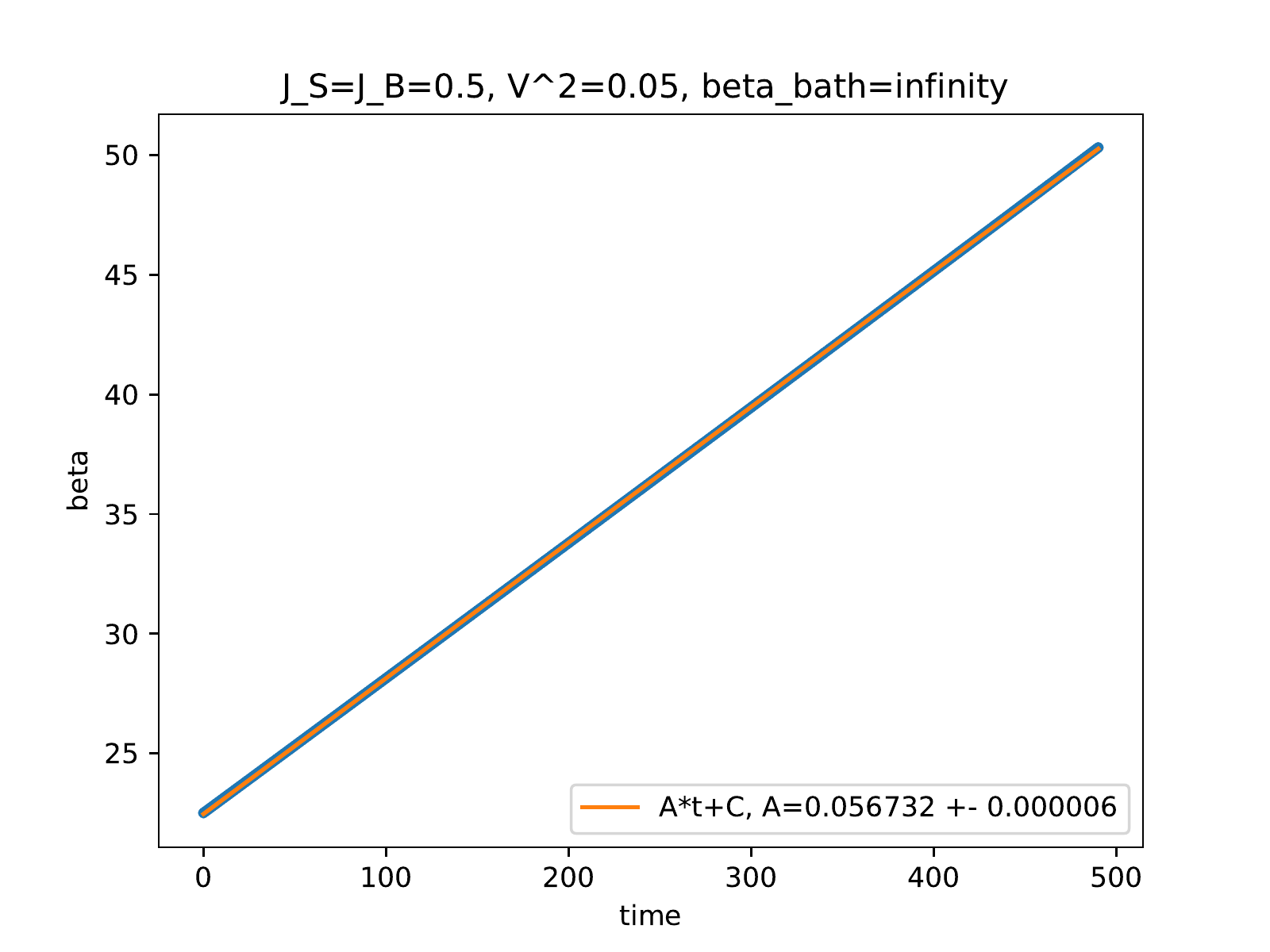}
\caption{Energy and $\beta$ for the irrelevant coupling $q_B=4, f_B=5, J_S=J_B=0.5$. The dense blue points are numerical data. The analytical answer for the slope is $0.063$ from Eq. (\ref{eq:lin}).}
\label{fig:lin}
\end{figure}

For an irrelevant deformation with $f_B=5$ we have
\beq
X_B = - G_B^5
\eeq
and the integral is
\beq
\label{int3}
\int_{-\infty}^{+\infty} dx \ \frac{1}{ (x - i \ep)^{5/2}} 
\frac{\cosh \pi  \l x - i \ep \r}{\sinh^{3/2} \pi  \l x  - i \ep \r} = -1.98.
\eeq
Hence, the temperature obeys
\beq
\beta' = 1.98 \frac{V^2 b^6 \sqrt{J_S}}{J_B^{5/2} \al_S 8 \sqrt{2\pi}}, \quad \text{irrelevant,} \ f_B=5.
\label{eq:lin}
\eeq
Again we have very good agreement with the numerics, see Figure \ref{fig:lin}.

\subsubsection{Irrelevant deformation: bath at finite temperature}

\begin{figure}[!ht]
\centering
\includegraphics[scale=0.5]{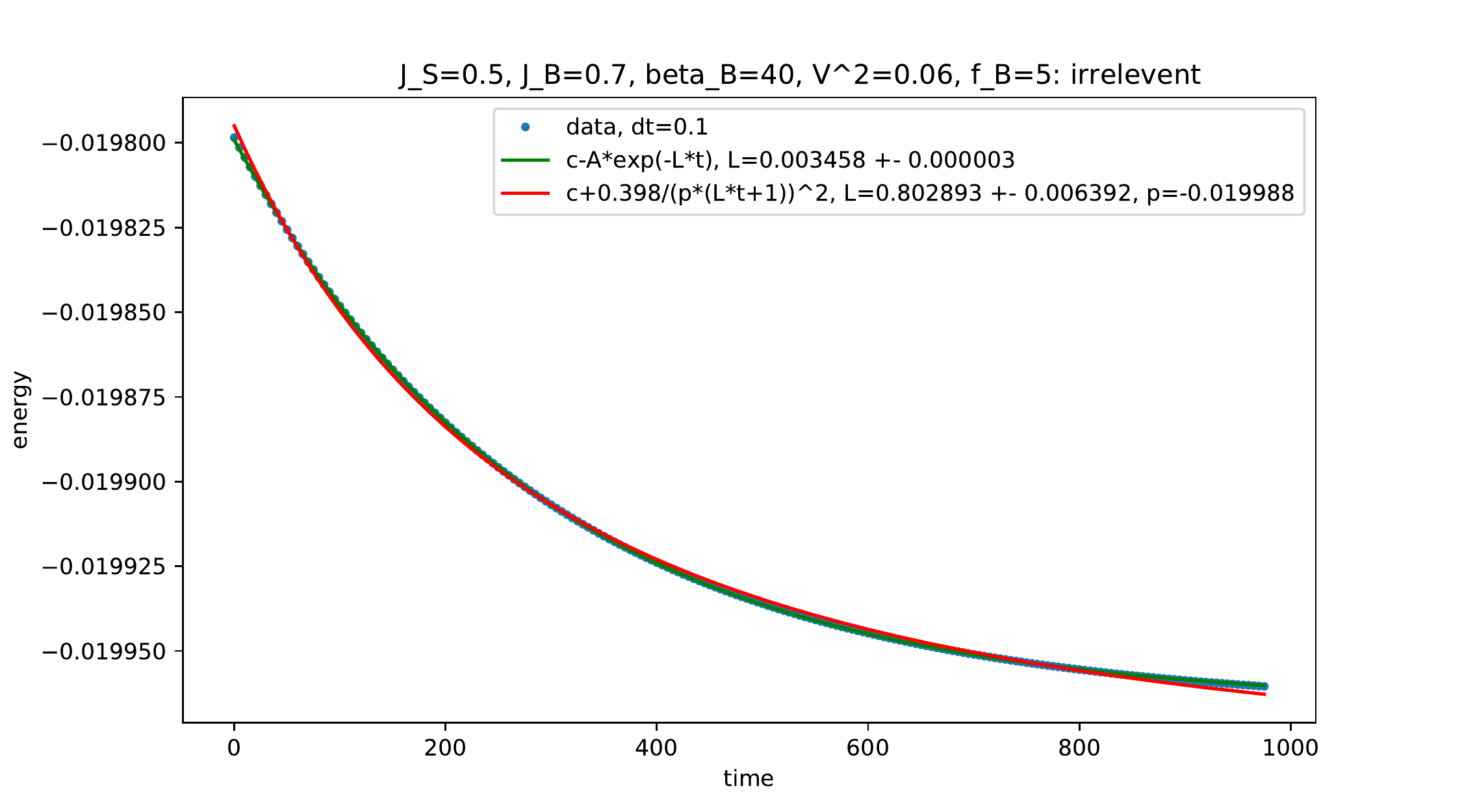}
\caption{Energy as a function of time for $\beta_{\rm init}=35$ and $\beta_B=40$. Only the late time behavior is shown. The green line is an exponential fit. For comparison we
also included a fit with $E \sim 1/t^2$. The analytical answer for the rate is $0.0034$ from Eq. (\ref{fin_p5}).}
\label{beta40_p5_jj}
\end{figure}

On physical grounds, we expect that if the system and the bath have close temperatures then the flux 
will be proportional to the temperature difference.
Indeed, if $\beta_B \sim \beta$ we again get exponential approach:
\begin{align}
\label{int4}
\int_{-\infty}^{+\infty} dx \ \frac{\pi^{5/2} \beta^{5/2}}{ \beta_B^{5/2} \sinh^{5/2} \l \pi (x-i \ep) \frac{\beta}{\beta_B} \r} 
\frac{\cosh \pi  \l x - i \ep \r}{\sinh^{3/2} \pi  \l x  - i \ep \r} = \\ \nonumber
= -\frac{5}{2} \pi^{7/2} \l \frac{\beta}{\beta_B} -1  \r \int_{-\infty}^{+\infty} dx \ \frac{ (x- i \ep) \cosh^2 \pi \l x - i \ep \r }{ \l \sinh^2 ( \pi (x-i \ep)) \r^{10/4} }  = \\ \nonumber
= 8.57 \frac{\beta-\beta_B}{\beta_B}.
\end{align}
Hence, the temperature obeys
\beq
\beta' = 8.57 (\beta_B-\beta) \frac{V^2 b^6 \sqrt{J_S}}{8 \sqrt{2\pi} J_B^{5/2} \beta_B}, \quad \text{irrelevant,} \ f_B=5.
\label{fin_p5}
\eeq
For $J_S=0.5, J_B=0.7$ the agreement is again very good as shown in Figure \ref{beta40_p5_jj}.

\subsection{Late time: approach to equilibrium and black hole evaporation}
\label{sec:total_evaporation}
We have seen that the late time approach to true equilibrium is in many cases exponential. This agrees with our physical expectations since the heat flux should be proportional to the temperature difference.

Obviously, this process of energy flow cannot last forever. At the very least, the temperature has
thermodynamic fluctuations,
\beq
\bra  (\Delta T)^2\ket = \frac{T^2}{C_v},
\eeq
where $C_v$ is the heat capacity, which is of order $N$ for SYK if the temperatures is not too low. These fluctuations imply that once the difference $T_S-T_{\rm bath}$ becomes of order  $\Delta T \sim 1/\sqrt{N}$, we have effectively reached true equilibrium. In the situations studied so far, this will take a time of order $\log N$.

However, one important point is the way $V^2$ scales with $N$. For an evaporating black hole in which the energy transfer is accomplished by a small number of light fields, the energy loss rate should be of order $N^0$ instead of order $N^1$. This can be modeled by taking $V^2$ to scale with $N$ as $N^{-1}$ instead of as $N^0$, e.g.
\beq
V^2 \sim \frac{V_0^2}{N},\ V_0^2 \sim N^0.
\eeq
Our analysis is still valid in this case, because we can use the classical Schwarzian description until $\beta \sim N$ and it is not important that the perturbation has $1/N$ suppression. Hence, the evaporation time becomes $ N \log N$.

However, this estimate is somewhat imprecise. As we just said, we can trust our classical computation in the previous subsection as long as $\beta \ll N/J_S$. Once $\beta$ becomes of order $N$ we have to quantize the Schwarzian. This is quite complicated given the non-local term (\ref{sch:int}) in the action. Hence, it appears challenging to derive an analogue of Eq. (\ref{eq:betap}) directly from the Schwarzian.

As an aside, notice that the problem does simplify when $V^2 \sim 1/N$ because when $\beta$ reaches $N/J_S$, the integration range in Eq. (\ref{flux:syk}) is already $N \log N$ (instead of $\log N$ when $V^2$ is order $N^0$), so we do not need to worry about the boundary term.

Fortunately, we do not need to carry out the full quantization procedure.\footnote{We are grateful to Alex Kamenev, Juan Maldacena and Luca Iliesiu for discussions about the following computation.} Recall that we derived Eq. (\ref{flux:syk}) for any Majorana system interacting with a Majorana bath. It happens that Eq. (\ref{eq:betap}) follows from it if we use the classical Schwarzian expression for the energy and the conformal approximation for the Green's functions. We will employ the same strategy in the quantum case. The exact expression for Schwarzian free energy is
\beq
F =  - \frac{2 \pi^2 \alpha_S}{\Jc_S \beta^2} + \frac{3}{2 N \beta} \log \beta,
\eeq
and the energy is 
\beq
\label{energy:low}
E = F + \beta \frac{d F}{d \beta} = \frac{2 \pi^2 \alpha_S}{ \Jc_S^2 \beta^2} + \frac{3}{2 N \beta}.
\eeq
In particular, when $\beta \sim N$ the last term dominates.

The behavior of SYK Green's function strongly depends on the relation between $t, \beta$ and $C$ where $C=N \alpha_S/\Jc_S$ is the coefficient in front of the Schwarzian term.
As long as $t,\beta \ll C$ we have the classical result:
\beq
G_S \sim \frac{1}{\sqrt{J \beta \sinh \l \frac{\pi t}{\beta} \r}}.
\eeq
Here and below we will suppress the numerical coefficients, but keep the factors of $\beta$ and $N$ explicit.
A generic answer for $G_S$ was obtained in \cite{mertens,kamenev}:
\begin{align}
\label{gs:full}
G_S \sim  \frac{1}{\sqrt{J_S C}} e^{- \frac{2 \pi C}{\beta} } \l \frac{\beta}{C}\r^{3/2}  \int_{-\infty}^{+\infty} d \mu(k_1) d\mu(k_2) \  \gamma(k_1,k_2)^2 \times \nonumber \\
\times \exp \l -\frac{1}{2c} 
\l -i t k_1^2 - (\beta - i t) k_2^2 - \ep k_1^2 -\ep k_2^2 \r \r ,
\\
\gamma(k_1,k_2)^2 = \frac{1}{\Gamma(1/2)} 
\Gamma(1/4 + i(k_1+k_2)) \Gamma(1/4 - i(k_1+k_2)) \times \nonumber \\ 
\times \Gamma(1/4 + i(k_1-k_2)) \Gamma(1/4 - i(k_1-k_2))\\
d\mu(k) = \sinh(2 \pi k) k dk .
\end{align}
Suppose that $\beta \gg C$ is large in the above expression. Then we can use the saddle-point approximation for $k_2$ with the following result:
\beq
G_S \sim  \frac{\beta^{3/2}}{\sqrt{J_S C}(t + i \beta)^{3/2}} 
\int_{-\infty}^{\infty} d \mu(k_1) \ e^{-\frac{1}{2c} 
\l -i t k_1^2 - \ep k_1^2 \r } \Gamma(1/4 + i k_1)^2 \Gamma(1/4-i k_1)^2.
\eeq
If $t \ll C$, then the integral is dominated by large $k_1$, so expanding the Gamma functions for large $k_1$ yields
\beq
\label{gs:early}
G_S \sim \frac{1}{\sqrt{J_S(t- i \ep)}},\ t \ll C \ll \beta,
\eeq
which is the expected result for zero-temperature case.
However, if $t \gg C$ we can use the saddle point approximation again, this time for $k_2$ \cite{mertens, kamenev}:
\beq
\label{gs:late}
G_S \sim \frac{N \beta^{3/2}}{J_S^{3/2} (t + i \beta)^{3/2} (t - i \ep)^{3/2}},\ t \gg C, \ \beta \gg N/J_S.
\eeq
One cross-check it that the expressions (\ref{gs:early}) and (\ref{gs:late}) coincide when $t \sim N/J_S$.

The last step before the actual calculation of the evaporation rate is the expression for $G_B$. As we mentioned before, the number of bath fermions $M$ must be much bigger than $N$. Here we assume $M$ is big enough to keep the bath classical even at large times $t$, so the Green's function is 
\beq
G_B^3 \sim \frac{1}{J_B^{3/2} (t- i \ep)^{3/2}}.
\eeq

All these pieces can now be assembled to compute the energy flux. Integrating by parts in Eq. (\ref{flux:syk}), we need to compute:
\beq
\label{preflux}
-\int_{-\infty}^{+\infty} dt \ \pr_t G_B^3 G_S \sim \frac{V_0^2}{N J_B^{3/2}} \int_{-\infty}^{+\infty} dt\ \frac{1}{(t- i \ep)^{5/2}} G_S(t).
\eeq
First of all, if we put $\beta=\infty$, the system's Green's
function (\ref{gs:late}) does not have singularities in the lower half-plane, so we can close the contour and get zero. This is expected: if both the system and the bath have zero temperature, then flux is zero.

To compute the integral at finite beta we consider the full integral representation (\ref{gs:full}). Notice that we can move the integral over $t$ in Eq. (\ref{preflux}) into the lower half-plane such that $t$ acquires constant imaginary
part of order $\beta$. In this case we can use the asymptotic formula in (\ref{gs:late}) for $G_S$. Hence, the flux is of order
\beq
 \sim \frac{V_0^2}{(J_S J_B)^{3/2}} \frac{1}{\beta^3}.
\eeq

Equating this to the loss of energy (\ref{energy:low}), we find only $\sqrt{t}$ behavior for $\beta$ instead of exponential growth:
\beq
\label{betap:late}
\beta' \sim \frac{V_0^2 N}{(J_S J_B)^{3/2}} \frac{1}{\beta},\ \beta \gg N/J_S,
\eeq
\beq
\beta(t) \sim \sqrt{ \frac{V_0^2 N t}{(J_S J_B)^{3/2}}}.
\eeq
As a check, note that for $\beta \ll N/J_S$ we had
\beq
\beta' \sim \frac{V_0^2}{N} \frac{\sqrt{J_S} \beta}{J_B^{3/2}},\ \beta \ll N/J_S
\eeq
from Eq. (\ref{betap:early}). Equations (\ref{betap:late}) and
(\ref{betap:early}) agree for $\beta \sim N/J_S$.

Thus, in the quantum regime there is a cross-over from exponential behavior, $\beta \sim e^t$, to power-law behavior, $\beta \sim t^{1/2}$.

\subsection{Checking the bound numerically}
\label{sec:syk_bound}

Having described all the parts of the curve analytically, let us discuss its precise form and check the proposed bound numerically. Our numerical setup is described in Appendix \ref{kb:numerics}. The main limitation comes from the fact that we cannot go to very low temperatures, because the Green's functions spread a lot. So we will limit ourselves to finite bath temperature. Also, we will study two kinds of interactions: marginal $f_S=1, f_B=3; f_S=2, f_B=2$ and irrelevant $f_S=1, f_B=5; f_S=5, f_B=1$.

At weak system-bath coupling, we do not expect a violation of the bound since we have a perturbative proof. However, at very strong coupling the final energy of the system is higher than the initial energy, because the interaction increases the ground state energy. Hence, something interesting might happen as we scan from weak coupling to strong coupling. 

\begin{figure}[!ht]
\centering
\includegraphics[scale=0.7]{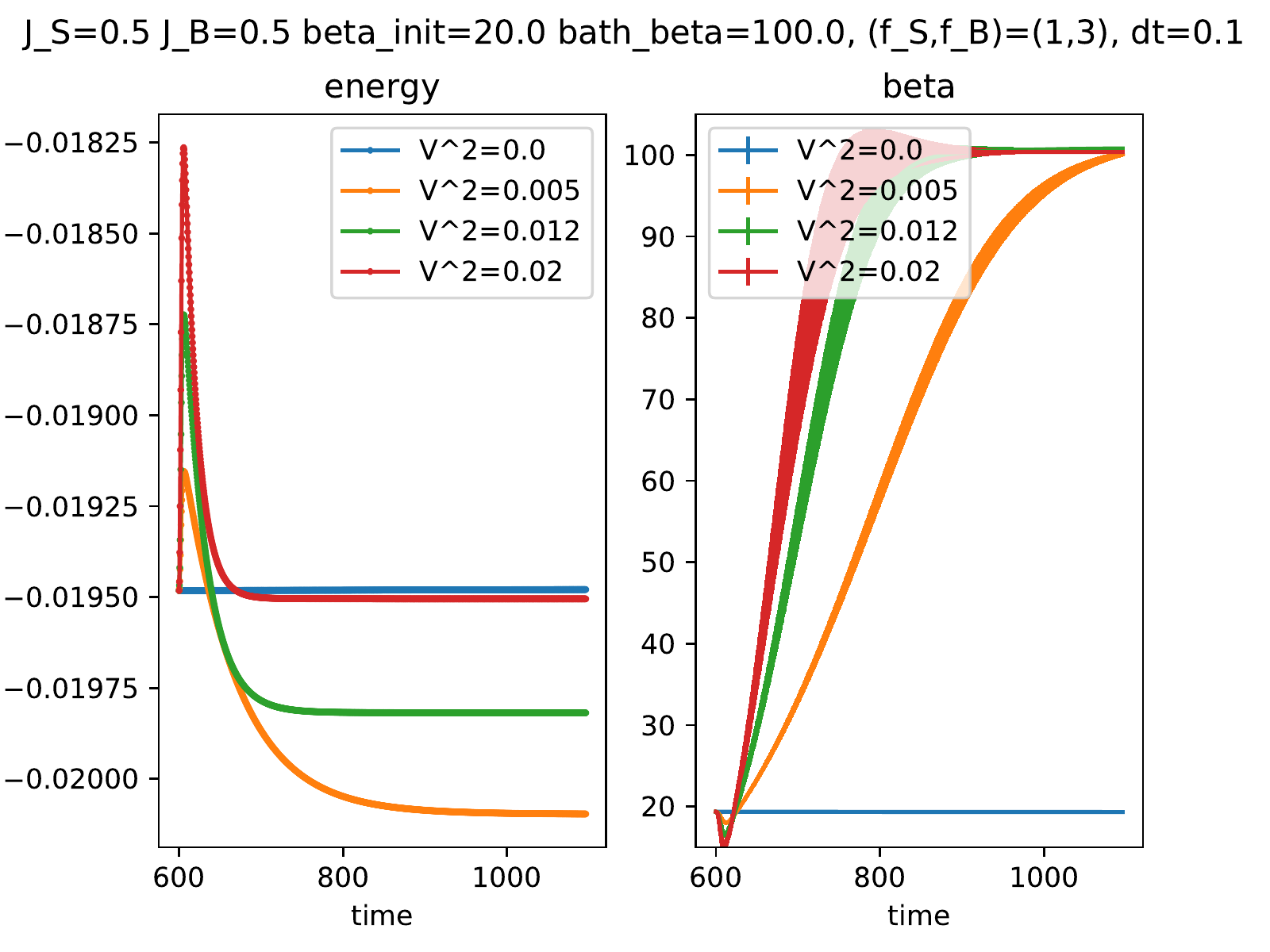}
\caption{Energy and beta as functions of time for the marginal deformation $(f_S,f_B)=(3,1)$ with $J_S=J_B=0.5, \beta_{\rm init}=20, \beta_{\rm bath}=100$. The thickness of the beta curve indicates the uncertainty in beta. }
\label{graph31}
\end{figure}

\begin{figure}[!ht]
\centering
\includegraphics[scale=0.7]{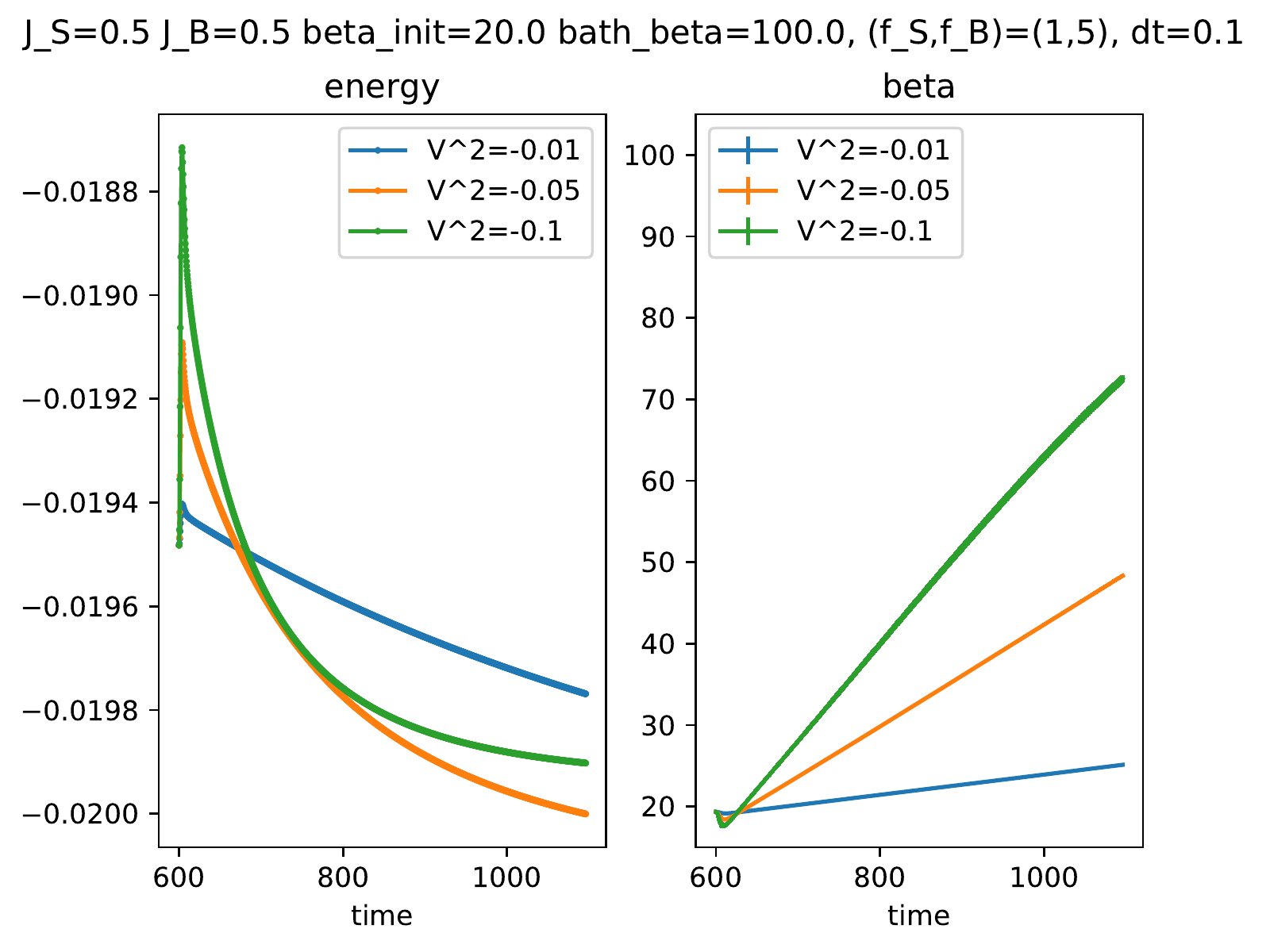}
\caption{Energy and beta as functions of time for the irrelevant deformation $(f_S,f_B)=(5,1)$ with $J_S=J_B=0.5, \beta_{\rm init}=20, \beta_{\rm bath}=100$. The thickness of the beta curve indicates the uncertainty in beta. }
\label{graph51}
\end{figure}

Our numerical results suggest that the integral in the bound is always bigger than zero. This is true even if we take $\beta(t)$ instead of the initial $\beta$. Our result are presented on Figures \ref{graph31}, \ref{graph51} and Tables \ref{table31}, \ref{table51}.   
The main source of error is the fact that the energy not conserved even for $V=0$ because of the discretization scheme, so we include the $V=0$ case for reference.

\begin{table}
\centering
\begin{tabular}[!ht]{|c|c|c|}\hline
$V^2$ & $F_{2/\beta_0}$ & $F_{2/\beta(t)}$ \\ \hline
  0.0 & $ \pm 10^{-7}$ & $ \pm 10^{-7}$   \\ \hline
  0.005 & $(2.05 \pm 0.01) \times10^{-4}$ & $(2.06 \pm 0.01) \times10^{-4}$ \\ \hline
  0.012 & $(4.68 \pm 0.01) \times10^{-4}$ &  $(4.73 \pm 0.01) \times10^{-4}$ \\ \hline
  0.02 & $(7.47 \pm 0.02 )\times10^{-4 }$ &  $(7.70 \pm 0.01)\times 10^{-4}$ \\ \hline
\end{tabular}
\caption{Results for $F_{2/\beta}$ for the marginal deformation $(f_S,f_B)=(3,1)$ with $J_S=J_B=0.5, \beta_{\rm init}=20, \beta_{\rm bath}=100$.  The errors were estimated by comparing the
results of $dt=0.1$ and $dt=0.05$.}
\label{table31}
\end{table}

\begin{table}
\centering
\begin{tabular}[!ht]{|c|c|c|}\hline
$V^2$ & $F_{2/\beta_0}$ & $F_{2/\beta(t)}$ \\ \hline
  0.01 & $ (5.13 \pm 0.03) \times 10^{-5}$ & $ (5.17 \pm 0.03) \times 10^{-5}$   \\ \hline
  0.05 & $(2.48 \pm 0.02) \times10^{-4}$ & $(2.51 \pm 0.02) \times10^{-4}$ \\ \hline
  0.1 & $(4.79 \pm 0.04) \times10^{-4}$ &  $(4.89 \pm 0.03) \times10^{-4}$ \\ \hline
\end{tabular}
\caption{Results for $F_{2/\beta}$ for the irrelevant deformation $(f_S,f_B)=(5,1)$ with $J_S=J_B=0.5, \beta_{\rm init}=20, \beta_{\rm bath}=100$. The errors were estimated by comparing the
results of $dt=0.1$ and $dt=0.05$.}
\label{table51}
\end{table}

\subsection{Comparison to exact finite $N$ calculations}
\label{sec:syk_ed}

Finally, we verify that the qualitative features of the energy curve persist at small $N$ via direct numerical integration of the Schrodinger equation. Because it enables us to access larger system sizes, we work with pure states instead of mixed states and integrate the full system-bath Schrodinger equation using a Krylov approach.

As above, the system is a $q_S$-SYK model with $N$ fermions while the bath is $q_B$-SYK model with $M$ fermions. The fermions are represented in terms of spins using a standard Jordan-Wigner construction. To prepare the inital state, we begin with a product state in the spin basis and evolve in imaginary time to produce:
\begin{equation}
    |\psi_{\text{ini}} \rangle \propto e^{- \beta_S H_S/2 - \beta_B H_B/2} |\text{product}\rangle.
\end{equation}
The coupling is then suddenly turned on at time $t=0$ and the full system-bath composite is evolved forward in time. The energy of the system as well as the system-bath entanglement are measured as a function of time.

\begin{figure}
    \centering
    \includegraphics[width=.9\textwidth]{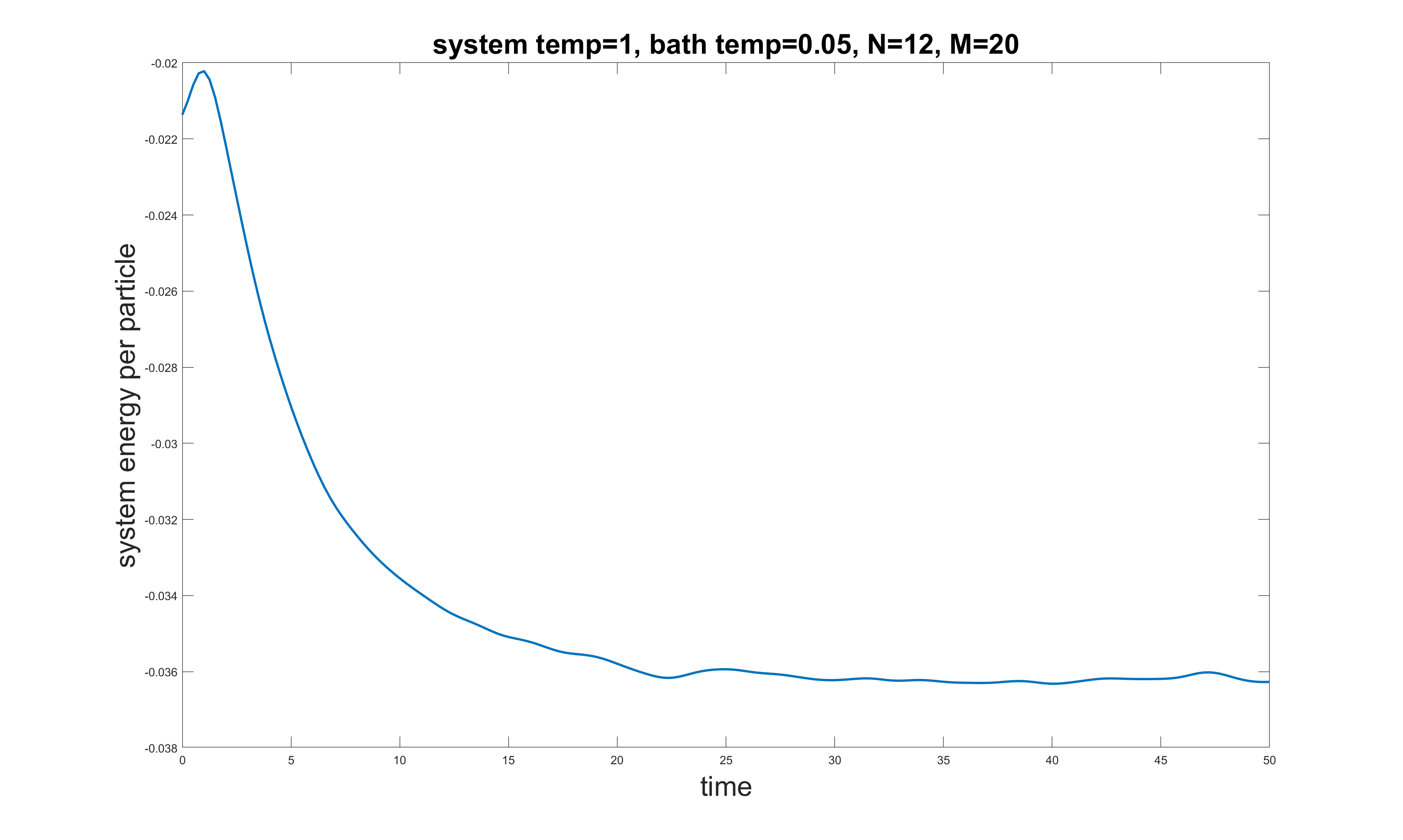}
    \caption{Energy curve for a finite size system-bath setup with $N=12$ and $M=20$. Other parameters are discussed in the main text.}
    \label{fig:energy_curve_ed}
\end{figure}

In Figure \ref{fig:energy_curve_ed} we show an example of the energy curve for $N=12$, $M=20$, $q_S=4$, $q_B=4$, $f_S=1$, $f_B=3$, and $g=.2$. The initial temperatures were $\beta_{S0} = 1$ and $\beta_{B0}=20$ in units where $J_S=J_B=1$. One clearly sees the initial energy bump, the subsequent cross-over to energy loss, the slow draining of energy into the bath, and a final approach to true equilibrium. Note that the final equilibrium system energy is modulated by finite size fluctuations in the time domain. The data shown constitute a single disorder sample with no disorder averaging. Also, the bound is satisfied for this example.

\section{Discussion}

Inspired by the problem of black hole evaporation, we studied in the detail the physics of thermalization for a system suddenly coupled to a bath. Our first key result is a positivity bound on the integrated energy flux. We proved it in general in perturbation theory and showed that it implied an instance of the ANEC. Our second key result is a detailed study of the thermalization dynamics for two coupled SYK clusters. In particular, at low energy we gave a thorough analytical discussion of the energy curve.

There are many directions for future work. One is to understand how far beyond perturbation theory our bound extends. In the SYK example, we found it to be quite robust. We suspect that quantum information ideas will be useful in this context, partly because the bulk interpretation of the bound in terms of the ANEC is associated with prohibiting unphysical communication between two entangled parties. There is also more to understand about the SYK case, for example, it may be possible to analytically solve the dynamical equations at large $q$.

More generally, it would be interesting to generalize our analysis to pure states, and to understand in detail the behavior of the entanglement entropy of various parts of the system. 
Finally, it is tempting to try to relate our rigorous Planckian bound on the energy curve to other more speculative Planckian bounds, for example, in transport physics. One idea for relating them is to the use the fact that dissipative transport generates heat, so perhaps this fact can be combined with some version of the setup we considered here?

\subsection*{Acknowledgments}
We would like to thank Juan~Maldacena for many comments and
discussions. It's a pleasure to thank Alexander~Abanov, Ksenia~Bulycheva, Yiming~Chen, Alexander~Gorsky, Alex~Kamenev, Dmitri~Kharzeev, Ho~Tat~Lam, Fedor~Popov, Jacobus~Verbaarschot, Douglas~Stanford, Zhenbin~Yang for helpful comments on this topic. A.A. is supported by funds from the Ministry of Presidential Affairs, UAE. BGS is supported by the Simons Foundation via the It From Qubit collaboration.

\appendix

\section{Numerical setup for KB equations}
\label{kb:numerics}
This appendix describes our approach to the numerical solution of the Kadanoff--Baym equations. Our strategy is based on previous work on SYK reported in \cite{quench1,quench2}. 

\begin{figure}[!ht]
\centering
\includegraphics[scale=0.7]{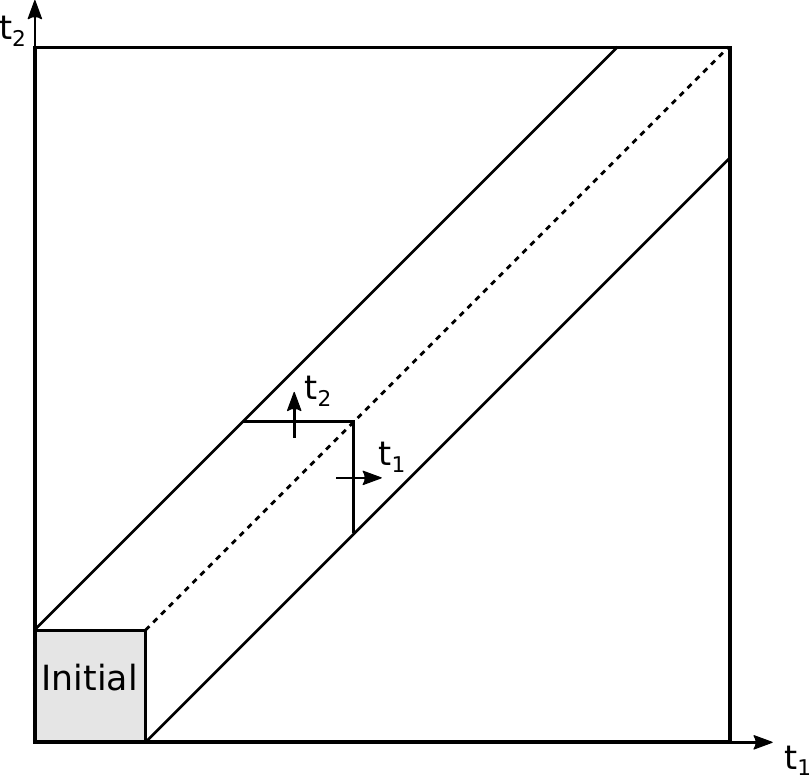}
\caption{Geometry of the $(t_1,t_2)$ plane. The initial Green's functions are placed inside the gray box.}
\label{grid}
\end{figure}

We use a uniform two-dimensional grid to approximate the $(t_1,t_2)$ plane. The grid spacing $dt$ plays the role of a UV cutoff and should be much smaller than $1/J_S$ and $1/J_B$. To fix the units of energy and time, we set $J=0.5$. In these units, we consider three grid spacings: $dt=0.2$, $dt=0.1$, and $dt=0.05$. The primary numerical cost arises from the grid size, as the overall size must be large to study low temperature effects. Typically, the Green's functions decay exponentially, so the calculation can be streamlined by restricting attention to a strip $|t_1-t_2| \lesssim c \beta_{\rm max}$ as shown in Figure \ref{grid}. All Green's functions are put to zero outside the strip. We take $\beta_{\rm max}$ to be the largest $\beta$ in the problem, typically the inverse bath temperature. In practice, $c$ is taken large enough to see converged results.

The initial Green's function is found by numerically solving the Lorentzian Schwinger--Dyson equation in equilibrium. To compute the integral in the KB equations we use the trapezoid method, and for the time propagation, we use a predictor-corrector scheme. Some care is needed when propagating along the diagonal. Fortunately, for Majorana fermions there is a simple relation:
\begin{equation}
G^>_S(t,t) = -\frac{i}{2}.
\end{equation}
However, for the Green's function obtained by numerically solving the DS equation the diagonal value is not exactly $-i/2$, so on a discrete lattice we just propagate this value:
\beq
G^>_S(j,j)=G^>_S(0,0).
\eeq

\begin{figure}[!ht]
\centering
\includegraphics[scale=0.5]{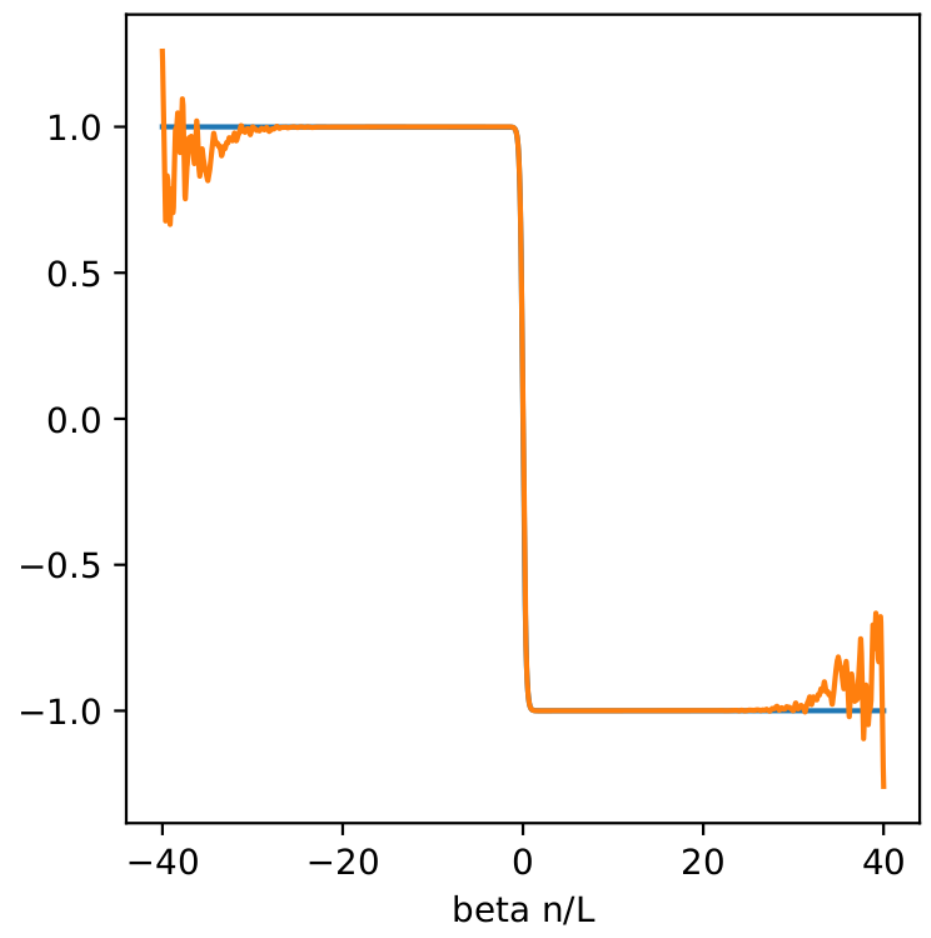}
\caption{(Orange) The left hand side of eq. (\ref{tanh}) for $\beta=80$, $dt=0.2$. The Green function
is localized on a strip of width $L=1600$. (Blue) $tanh$ function. They coincide up to frequencies corresponding
to the timescale $\sim \beta/30 \sim 2.6$, which is comparable to the timestep $0.2$. }
\label{fig:tanh}
\end{figure}

The integral in the bound (\ref{the_bound}) is also computed using the trapezoid rule and the energy time derivative is discretized in a simple way: $(E_{j+1}-E_j)/dt$.
To estimate the error the integral is computed for different time steps and without the coupling to the bath. Note also that we can not really integrate all the way to infinity. In order not to rely on any extrapolations, we use a crude upper-bound for the error obtained from integration over a finite interval. Obviously, since the flux is decreasing and beta in increasing to $\beta_b$ we have
the following inequality:
\beq
\int_{t_f}^{\infty} dt \ E_S' e^{-\kappa t/\beta(t)} \le E_S'(t_f) \int_{t_f}^{+\infty} \ dt e^{-\kappa t/\beta_b} = \frac{1}{\kappa} E_S'(t_f) \beta_b e^{-\kappa t_f/\beta_b}
= I^\text{err}_\kappa.
\eeq
Finally, there is a question of how to define the temperature at particular time $t$ in non-equilibrium situation? There are two possibilities. We can consider the
``diagonal slice'' $G_\text{eq}^>(t;\delta)$:
\begin{equation}
G_{\text{eq},t}^>(\delta)=G^>_S(t-\delta,t+\delta),
\label{}
\end{equation}
treat it as a two-point function in the equilibrium, and find the temperature using the 
fluctuation-dissipation theorem (FDT):
\beq
\frac{\Im(G^>_{\text{eq},t}(\omega)+G^<_{\text{eq},t}(\omega))}{(-2) \Im G^R_{\text{eq},t}(\omega)} = - \tanh \frac{\beta(t) \omega}{2}.
\label{tanh}
\eeq
However, this choice does not respect causality in time. Another choice is the ``corner slice'': $G^>_\text{eq}(t,\delta)$:
\begin{equation}
G^>_{\text{eq},t}(\delta)=\theta(\delta) G^>_S(t-\delta,t)+ \theta(-\delta) G^>_S(t,t+\delta).
\end{equation}
This choice respects causality and this corner Green's function enters in the definition of energy (\ref{energy:lorentz}). Therefore we will adopt the corner definition. Unfortunately, the FDT holds for low frequencies only, since large frequencies are affected by the size of the
discretization step. However in all our setups the relation (\ref{tanh}) holds for low frequencies up to the frequencies
of order of the discretization step $1/dt$.
see Figure \ref{fig:tanh} for a typical behavior.

\section{Energy flux from KB equations}
\label{flux_kb}

Consider the case $q=4$ and one system fermion in the interaction ($f_S=1$). The generalization to other cases is straightforward. For clarity, we denote the system's ``greater'' Green's function $G_>$ simply by $G_S$ and bath's ``greater'' function as $G_B$.
The energy is given by:
\beq
E_S(t_1) =-i  \frac{J_S^2}{4} \int_{-\infty}^t dt_2 \ \l  G_S(t_1,t_2)^4 - G_S(t_2,t_1)^4 \r.
\eeq
Our approach is to differentiate it and use the KB equations. We will assume that the coupling to the bath is switched on at time $t=0$:
\begin{align}
\frac{dE_S(t_1)}{dt_1} =   V^2 J_S^2 \int_{-\infty}^{t_1} dt_2 \int_{0}^{t_1} dt_3 \l G_B(t_1,t_3)^{f_B} + G_B(t_3,t_1)^{f_B}  \r \Big( G_S(t_1,t_2)^3 G_S(t_3,t_2) -  \nonumber \\ 
G_S(t_2,t_1)^3 G_S(t_2,t_3)  \Big) +  \nonumber \\
 V^2 J_S^2 \int_{-\infty}^{t_1} dt_2 \int_{0}^{t_2} dt_3 \l G_S(t_3,t_2) + G_S(t_2,t_3)  \r \Big( - G_S(t_1,t_2)^3 G_B(t_1,t_3)^{f_B} +  \nonumber \\
 G_S(t_2,t_1)^3 G_B(t_3,t_1)^{f_B}  \Big).
\end{align}

This is the leading term in $V$. If we trust this expansion, then we can use the initial $G_S$ to find the flux.

We can exchange the integration order in the second term:
\beq
\int_{-\infty}^{t_1} dt_2 \int_0^{t_2} dt_3 \ra \int_0^{t_1} dt_3 \int_{t_3}^{t_1} dt_2 \ra
\int_0^{t_1} dt_3 \l \int_{-\infty}^{t_1} dt_2 - \int_{-\infty}^{t_3} dt_2 \r.
\eeq
After that, we use the equilibrium Dyson--Schwinger equation for $G_S$ to convert the convolution
over $t_2$ into a time derivative, arriving at
\beq
E_S' = i V^2 \int_{-t}^t du G_B(u-i \ep)^{f_B} \pr_u G_S(u- i \ep).
\eeq

\section{Locating the peak}
\label{app:peak}

Suppose that the bath is ``fast'': $J_B \beta \ll 1$. If we assume that $t_{\rm peak} J_B \ll 1$, then $G_B$ varies slowly and we can Taylor expand it around $u=0$ to greatly simplify the result. In the first part of Eq. (\ref{eq:part_int_im}) we can simply put $G_B^{f_B}=-i/2$, whereas in the 
integral we put $\pr_u G_B^{f_B} = -3 J_B /2$. Up to an overall coefficient the flux is
\beq
\Im \l \frac{i \sqrt{i}}{\sqrt{\sinh \l \frac{\pi u}{\beta}\r}} + 
\frac{3 \beta J_B}{\pi} \EllipticF \l \frac{\pi (\beta-2 i u)}{4\beta} ,2  \r  \r
\eeq
Assuming $u /\beta \gg 1$ we can approximate the elliptic function by a constant:
\beq
\EllipticF \l \frac{\pi (\beta-2 i u)}{4\beta} ,2  \r \approx -i \EllipticK(-1).
\eeq

Now it is easy to balance the two terms to estimate
\beq
t_{\rm peak} \sim \beta \log \l \frac{1}{J_B \beta} \r, \quad J_B \beta \ll 1.
\eeq
Note that both assumptions, $t_{\rm peak}/\beta \gg 1$ and $\ t_{\rm peak} J_B \ll 1$, are satisfied.

In the opposite regime, $J_B \beta \ll 1$, we assume that $t_{\rm peak}/\beta \ll 1$. Then we can expand $G_S$:
\beq
G_S  = \frac{b \sqrt{i \pi}}{\sqrt{J_S \beta}} \l \frac{\sqrt{\beta}}{\sqrt{\pi u}} -
\frac{1}{12} \l \frac{\pi u}{\beta } \r^{3/2} \r.
\eeq
Now the integral in Eq. (\ref{eq:part_int_im}) can be computed analytically to give a lengthy expression
with rational functions and a logarithm. However, if we assume that $t_{\rm peak} J_B \gg 1$ and expand in large time $u$, then we find that the
leading imaginary contributions are
\beq
2 i \beta^2 J_B^2 - u^3 J_B^2 \pi^2 \log \l 4 i J_B (u - i \ep) \r.
\eeq
Thus, we have the estimate:
\beq
t_{\rm peak} \sim \frac{\beta}{\l \beta J_B \r^{1/3}} , \quad J_B \beta \gg 1.
\eeq
Again, our two assumptions, $t_{\rm peak} J_B \gg 1$ and $t_{\rm peak}/\beta \ll 1$, are satisfied.

\section{Equation of motion in Schwarzian}
\label{sch:eom}

It is convenient to first compute the variation with respect to a general reparametrization $t(u)$, then later to plug in the thermal solution,
\beq
t[u] = \tanh \l \frac{\pi u}{\beta} \r.
\eeq

As is well-known, the variation of the Schwarzian is equal to minus the time derivative of the Schwarzian. After we vary with respect to $t_+ - t_-$ and put $t_+ = t_-$, the leading term is
\beq
\pr_t \l \Sch(t[u],u) \r =  \frac{\beta}{\pi}
\cosh^2 \frac{\pi u}{\beta} \pr_u \l \Sch(t[u],u) \r 
=\frac{\beta}{\pi}
\cosh^2 \frac{\pi u}{\beta} \pr_u \l \frac{2 \pi^2}{\beta^2} \r.
\label{lag_kin}
\eeq

Now we deal with the interaction term. One has to vary
\beq
\int_\Cc du_1 du_2 \ X_B(u_1-u_2)
\l \frac{t_1' t_2'}{(t_1-t_2)^2} \r^{1/4}.
\eeq
Since $V^2$ is already small, after taking the variation we can plug in the thermal solution.
Another reason for this is that the integral is
dominated by $u_{12} < \beta$, whereas $\beta[u]$ changes on scales much bigger than $\beta$. 

Variation with respect to the $t_1$ yields
\begin{align}
- \int_\Cc du_2 \ 
\l X_B(u_1-u_2) \frac{(t_1' t_2')^{1/4}}{2(t_1-t_2)^{3/2}} +
\pr_{u_1} \l X_B(u_1-u_2) \frac{(t_1' t_2')^{1/4}}{4 t_1' (t_1-t_2)^{1/2}} \r\r.
\end{align}
Recall that the combination
\beq
\frac{(t_1' t_2')^{1/4}}{(t_1-t_2)^{1/2}} = \l \frac{\pi/\beta}{\sinh \l \frac{\pi(u_1-u_2)}{\beta}\r} \r^{1/2}
\eeq
is proportional to $G_S(u_1-u_2)$, so it is a function only of the difference $u_1-u_2$. Therefore, if $\pr_{u_1}$ does not act on $1/t_1$ it can be transformed into $\pr_{u_2}$ to give a total derivative.

After taking the derivative we will have
\begin{align}
-\int_\Cc du_2 \ X_B(u_1-u_2)
\frac{\cosh \frac{\pi u_1}{\beta}}{2 \sinh^{3/2} \l \frac{\pi (u_1-u_2)}{\beta} \r} \l \cosh \frac{\pi u_1}{\beta} + \sinh \frac{\pi u_1}{\beta} \sinh \frac{\pi (u_1-u_2)}{\beta} \r ,
\end{align}
which is equal to
\begin{align}
-\int_\Cc du_2 \ X_B(u_1-u_2)
\frac{\cosh^2 \frac{\pi u_1}{\beta}}{2\sinh^{3/2} \l \frac{\pi (u_1-u_2)}{\beta} \r}.
\end{align}
Hence, in the end we have:
\begin{align}
-2 \times \cosh \l \frac{\pi u_1}{\beta} \r^2 \int_\Cc du_2 \   X_B(u_1-u_2)
\frac{
\cosh  \frac{\pi}{\beta} \l u_1- u_2 \r}{2 \sinh^{3/2} \frac{\pi}{\beta} \l u_1-u_2 \r},
\end{align}
where an extra factor of 2 came from a similar variation with respect to the $t_2$.

Now we need to remember that we are working on the Keldysh contour, and the variation is over $t_+ - t_-$. This gives four pieces:
\begin{itemize}
\item $u_{1,+}$, $u_{2,+}$,
\item $u_{1,-}$, $u_{2,-}$,
\item $u_{1,+}$, $u_{2,-}$,
\item $u_{1,-}$, $u_{2,+}$.
\end{itemize}
These integrals combine into twice the integral
over the Wightman functions:
\beq
2 \times \int_{-\infty}^{+\infty} du_2.
\eeq
The last step is to change variables: $u_2 \ra x \beta + u_1$ and combine this contribution with that of the kinetic term (\ref{lag_kin}):
\begin{align}
\frac{8 \pi^2 \al_S}{\Jc_S \beta^3} \beta' = \frac{i \sqrt{i} b V^2 \pi^{3/2}}{  (J_S)^{1/2} \beta^{1/2}} \int_{-\infty}^{+\infty} dx \ X_B(\beta(u- i \ep))
\frac{\cosh \pi  \l x - i \ep \r}{\sinh^{3/2} \pi  \l x  - i \ep \r}.
\end{align}

Note that the factor of $\cosh^2 (\pi u_1/\beta)$ has cancelled out, meaning that the ansatz with slowly-varying beta is actually consistent with the equations of motion.

\section{Bounds on energy flow}
\label{app:bound}

\subsection{Perturbative energy flow calculation for bosonic coupling}

Consider a system initially in a thermal state of the form
\begin{equation}
    \rho_0 = \frac{e^{-\beta_S H_S - \beta H_B}}{Z_S Z_B}
\end{equation}
where the initial system Hamiltonian is $H_S$ and the bath Hamiltonian is $H_B$. At time zero, a system bath coupling $g H_{SB}$ is turned on, at which point the full Hamiltonian is
\begin{equation}
    H = H_0 + g H_{SB}
\end{equation}
where $H_0 = H_S + H_B$. 

The rate of the change of the system energy as a function of time is
\begin{equation}
    E'_S = \text{tr}\left(\rho_0 e^{i H t} [i g H_{SB},H_S] e^{-i H t}\right).
\end{equation}
This equation follows from the fact that $[H,H_S] = g [H_{SB},H_S]$. Let us assume that the system-bath coupling is of the form
\begin{equation}
    H_{SB} = O_S O_B,
\end{equation}
noting that the most general coupling is a sum of such terms. Then the rate of energy change is
\begin{equation}
    E'_S = \langle e^{i H t} i g [O_S,H_S]O_B e^{- i H t} \rangle_0.
\end{equation}

To work perturbatively in $g$, we move to the interaction picture, defining
\begin{equation}
    e^{-i H t} = e^{-i H_0 t} U.
\end{equation}
To zeroth order in $g$, $U$ is simply the identity, in which case $E'=0$ as follows form the thermality of the initial state.

To first order in $g$, $U$ is given by
\begin{equation}
    U = 1 -i g \int_0^t dt' H_{SB}(t')+\cdots
\end{equation}
where $H_{SB}(t')$ denotes the Heisenberg operator with respect to $H_0$ at time $t'$. The rate of energy change to second order in $g$ is 
\begin{equation}
    E'_S = (ig)^2 \int_0^t dt' \langle [O_S(t')O_B(t'),[O_S,H_S](t) O_B(t) ] \rangle_0.
\end{equation}

From the equation of motion
\begin{equation}
    [O_S,H_S](t) = i \partial_t O_S(t), 
\end{equation}
it follows that
\begin{equation}
    E'_S = i  g^2 \int_0^t dt' \langle [\partial_t O_S(t)  O_B(t), O_S(t')O_B(t') ] \rangle_0.
\end{equation}
Note that the commutator has also been reversed, hence the extra minus sign. There are two terms from the commutator,
\begin{align}
    [\partial_t O_S(t)  O_B(t),O_S(t')O_B(t'), ] =\, &  \partial_t O_S(t) O_S(t') [O_B(t),O_B(t')] \nonumber \\
    &+ [\partial_t O_S(t), O_S(t')] O_B(t') O_B(t).
\end{align}
Since the initial state factorizes, it follows that the energy rate of change can be written as a sum of products of system and bath correlators; these are defined as
\begin{equation}
    X_{S/B}=\langle O_{S/B}(t) O_{S/B}(t') \rangle_0
\end{equation}
and
\begin{equation}
    X^R_{S/B}=-i \langle [O_{S/B}(t), O_{S/B}(t')] \rangle_0.
\end{equation}
Since the $O$ operators are Hermitian, it follows that $\langle O(t') O(t)\rangle = \langle O(t) O(t')\rangle^*$. Then the rate of energy change is
\begin{equation}
    E'_S = - g^2 \int_0^t dt' \left\{ \partial_t X_S(t-t') X_B^R(t-t') + \partial_t X_S^R(t-t') X^*_B(t-t') \right\}.
\end{equation}
Note that technically the time derivative acts on both $\theta(t)$ and $O_S(t)$ in $X^R_S$, but this doesn't matter because $O_S$ commutes with itself. We have used the fact that the initial state is thermal to conclude that the dependence on $t,t'$ reduces to a dependence on $t-t'$ only.

It is useful to rewrite the two terms in $E'_S$ using a spectral representation. The first term is
\begin{equation}
    \partial_t X_S(t-t') X_B^R(t-t') = \int \frac{d\omega}{2\pi} \frac{d\omega'}{2\pi} \frac{d\nu}{2\pi} e^{-i (\omega+\omega')(t-t')}\frac{-i \omega A_{S+}(\omega) A_{B}(\nu)}{\omega'+i0^+ - \nu}.
\end{equation}
The second term is
\begin{equation}
    \partial_t X^R_S(t-t') X^*_B(t-t') = \int \frac{d\omega}{2\pi} \frac{d\omega'}{2\pi} \frac{d\nu}{2\pi} e^{-i (\omega-\omega')(t-t')}\frac{-i \omega A_{S}(\nu) A_{B+}(\omega')}{\omega +i0^+ - \nu} .
\end{equation}
The integral over $t'$ in the expression for $E'_S$ can be done to yield
\begin{equation}
    \int_0^t dt' e^{-i (\omega+\omega')(t-t')} = \frac{1 - e^{-i(\omega+\omega')t}}{i(\omega+\omega')}
\end{equation}
and
\begin{equation}
    \int_0^t dt' e^{-i (\omega-\omega')(t-t')} = \frac{1 - e^{-i(\omega-\omega')t}}{i(\omega-\omega')}.
\end{equation}
With a view towards the desired inequality, one can integrate these expressions against $e^{-\kappa t}$ for arbitrary $\kappa$. The result is
\begin{equation}
    \int_0^\infty dt e^{-\kappa t} \frac{1 - e^{-i(\omega+\omega')t}}{i(\omega+\omega')} = \frac{1}{\kappa [\kappa + i(\omega+\omega')]}
\end{equation}
and
\begin{equation}
    \int_0^\infty dt e^{-\kappa t} \frac{1 - e^{-i(\omega-\omega')t}}{i(\omega-\omega')} = \frac{1}{\kappa [\kappa + i(\omega-\omega')]}.
\end{equation}

In the first term the $\omega'$ integral can be carried out by contour, similarly for the $\omega$ integral in the second term (because these frequencies do not appear in the spectral functions). Closing in the upper half plane gives
\begin{equation}
    \int \frac{d\omega'}{2\pi} \frac{1}{\kappa [\kappa + i(\omega+\omega')]} \frac{1}{\omega'+ i 0^+-\nu} = -\frac{1}{\kappa ( \nu + \omega - i \kappa)}
\end{equation}
and
\begin{equation}
     \int \frac{d\omega}{2\pi} \frac{1}{\kappa [\kappa + i(\omega-\omega')]} \frac{\omega}{\omega+ i 0^+-\nu} = \frac{-\omega' - \nu -i \kappa}{2\kappa (- \omega' + \nu - i \kappa)}.
\end{equation}

Adding back all the factors, the first term becomes
\begin{equation}
    -g^2 \int \frac{d\omega}{2\pi}\frac{d\nu}{2\pi} \frac{-i \omega A_{S+}(\omega) A_{B}(\nu)}{-\kappa[\omega+\nu - i\kappa]}
\end{equation}
and the second term is
\begin{equation}
    -g^2 \int \frac{d\omega'}{2\pi}\frac{d\nu}{2\pi}\frac{-i (-\omega' -\nu -i \kappa) A_{S}(\nu) A_{B+}(\omega')}{2 \kappa (-\omega'+ \nu - i \kappa)}.
\end{equation}
To compare the two terms, we relabel variables in both terms so that $\omega$ appears in $A_{S}$ and $\omega'$ appears in $A_{B}$. The full expression for the integrated energy rate of change is thus
\begin{equation}
    F_\kappa = i \frac{g^2}{\kappa} \int \frac{d\omega}{2\pi}\frac{d\omega'}{2\pi} \left[ -  \frac{\omega A_{S+}(\omega) A_{B}(\omega')}{\omega + \omega' - i \kappa} -  \frac{(\omega'+\omega +i \kappa) A_S(\omega) A_{B+}(\omega')}{2(-\omega'+\omega-i\kappa)}\right].
\end{equation}

The useful identity $A_+(-\omega)=A_-(\omega)$ gives
\begin{equation}
    \int d\omega A(\omega) f(\omega) = \int d\omega A_+(\omega) [f(\omega)- f(-\omega)].
\end{equation}
Applied to the first term ($\omega'$ integral), an equivalent integrand is
\begin{equation}
    A_{S+} A_{B+} \left(- \frac{\omega}{\omega+\omega'-i \kappa} + \frac{\omega}{\omega-\omega'-i \kappa}\right).
\end{equation}
Applied to the second term ($\omega$ integral), an equivalent integrand is
\begin{equation}
    -\frac{A_{S+}(\omega) A_{B+}(\omega')}{2} \left(\frac{\omega'+\omega + i \kappa}{-\omega' + \omega - i \kappa} - \frac{\omega'-\omega + i \kappa}{-\omega' -\omega - i \kappa}\right)
\end{equation}
or
\begin{equation}
    A_{S+}(\omega) A_{B+}(\omega')\left( \frac{\omega}{-\omega+\omega'+i\kappa} + \frac{\omega}{\omega+\omega'+i\kappa} \right) .
\end{equation}
The terms may be recombined to give 
\begin{equation}
     A_{S+}(\omega) A_{B+}(\omega') \left(- \frac{\omega}{\omega+\omega'-i \kappa}  + \frac{\omega}{\omega+\omega'+i\kappa} \right),
\end{equation}
thanks to a cancellation of two terms. The real part is then simply zero while the imaginary part is
\begin{equation}
    \Im :- \frac{2 \omega \kappa}{(\omega+\omega')^2 + \kappa^2}.
\end{equation}
Combined the imaginary overall prefactor, it follows that the integrated rate of change is
\begin{equation}
  F_\kappa = \int dt e^{-\kappa t} E'_S =  2 g^2 \int \frac{d\omega}{2\pi}\frac{d\omega'}{2\pi} \frac{\omega A_{S+}(\omega) A_{B+}(\omega')}{(\omega+\omega')^2 + \kappa^2}.
\end{equation}

The integrated flux simplifies in various limits. For example, $\kappa \rightarrow \infty$ corresponds to the short time limit. The integrated flux obeys
\begin{equation}
    F_{\kappa \rightarrow \infty} \rightarrow 2 \frac{g^2}{\kappa^2}  \int \frac{d\omega}{2\pi}\frac{d\omega'}{2\pi} \omega A_{S+}(\omega) A_{B+}(\omega').
\end{equation}
Using 
\begin{equation}
    \int d\omega \omega A_+(\omega) = \frac{1}{2} \int d\omega \omega A(\omega) = \int_{\omega \geq 0} d\omega \omega A(\omega) \geq 0,
\end{equation}
it follows that 
\begin{equation}
    F_{\kappa \rightarrow \infty} \geq 0
\end{equation}
in agreement with Almheiri's lemma.

The limit $\kappa \rightarrow 0$ corresponds to the long time limit in which case the flux is dominated by the late time value. The integral over $\omega$ can be done by replacing the denominator with a delta function of $\omega + \omega'$ times $\pi/\kappa$,
\begin{equation}
    F_{\kappa \rightarrow 0} \rightarrow - \frac{ g^2}{\kappa}\int \frac{d\omega'}{2\pi} \omega' A_{S+}(-\omega') A_{B+}(\omega').
\end{equation}
Relabeling $\omega'$ as $\omega$, the integral can be written
\begin{equation}
    F_{\kappa \rightarrow 0} \rightarrow - \frac{ g^2}{\kappa}\int \frac{d\omega}{2\pi} \omega \frac{A_S(\omega) A_B(\omega)}{(e^{\beta_S \omega}-1)(1 - e^{-\beta_B \omega})}.
\end{equation}
This form is convenient since $A(\omega)$ is antisymmetric and so the product of two is symmetric. The integral can also be written
\begin{equation}
    F_{\kappa \rightarrow 0} \rightarrow - \frac{ g^2}{\kappa}\int \frac{d\omega}{2\pi} \omega \frac{e^{(\beta_B - \beta_S)\omega/2} A_S(\omega) A_B(\omega)}{4 \sinh \frac{\beta_S \omega}{2} \sinh \frac{\beta_B \omega}{2}}.
\end{equation}
Using the symmetry of the integrand, this can be written once more as
\begin{equation}
    F_{\kappa \rightarrow 0} \rightarrow - \frac{ g^2}{\kappa}\int_{\omega \geq 0} \frac{d\omega}{2\pi} \omega \frac{\sinh \frac{(\beta_B - \beta_S)\omega}{2} A_S(\omega) A_B(\omega)}{2 \sinh \frac{\beta_S \omega}{2} \sinh \frac{\beta_B \omega}{2}}.
\end{equation}
This form is nice because it makes it clear that the energy flow is negative or positive depending only on whether $\beta_B - \beta_S$ is positive or negative.

\subsection{Review of spectral representation}

Here we briefly review the spectral representation used above. Consider a Hermitian operator $O$ in a system with Hamiltonian $H$ in a thermal state at temperature $T = 1/\beta$. The two correlators of interest are
\begin{equation}
    X=\langle O(t) O(0)\rangle
\end{equation}
and
\begin{equation}
    X^R(t)= -i \theta(t) \langle [O(t),O(0)]\rangle.
\end{equation}

Both correlators have an expansion in terms of energy eigenstates. These are
\begin{equation}
    X(t) = \sum_{n,m} p_n |\langle n|O |m\rangle|^2 e^{i (E_n-E_m)t}
\end{equation}
and
\begin{equation}
    X^R(t) = -i \theta(t) \sum_{n,m} p_n |\langle n|O |m\rangle|^2\left[ e^{i (E_n-E_m)t} -  e^{-i (E_n-E_m)t} \right].
\end{equation}
The Fourier transforms are
\begin{equation}
    X(\omega)=\int_{-\infty}^\infty dt e^{i \omega t} X(t) = \sum_{n,m} \sum_{n,m} p_n |\langle n|O |m\rangle|^2 2\pi \delta(\omega-(E_m-E_n))
\end{equation}
and
\begin{equation}
    X^R(\omega)= \sum_{n,m} \sum_{n,m} p_n |\langle n|O |m\rangle|^2 \left(\frac{1}{\omega+i 0^+ - (E_m-E_n)} - \frac{1}{\omega+i 0^+ + (E_m-E_n)} \right).
\end{equation}

The spectral function is defined by the equation
\begin{equation}
    X^R(\omega) = \int \frac{d\nu}{2\pi} \frac{A(\nu)}{\omega+i 0^+ - \nu},
\end{equation}
from which it follows that
\begin{equation}
    A(\nu)= A_+(\nu)-A_-(\nu)
\end{equation}
with
\begin{equation}
    A_{\pm}(\nu)=\sum_{n,m}p_n |\langle n | O |m \rangle|^2 2 \pi \delta(\nu \mp (E_m-E_n)).
\end{equation}
Exchaning $n$ and $m$ in the definition of $A_-$ shows that
\begin{equation}
    A_- = \sum_{n,m}p_m |\langle n | O |m \rangle|^2 2 \pi \delta(\nu - (E_m-E_n)),
\end{equation}
and using $E_m = E_n+\nu$ plus the explicit form of $p_m$, it follows that
\begin{equation}
    A_- = e^{-\beta \nu } A_+
\end{equation}
and that
\begin{equation}
    A(\nu)=(1-e^{-\beta \nu}) A_+(\nu).
\end{equation}
We also see that $X(\omega)$ obeys
\begin{equation}
    X(\omega) = A_+(\omega).
\end{equation}

\subsection{General argument for perturbative bound}

Once again, the integrated flux is
\begin{equation}
    F_\kappa = 2 g^2\int \frac{d\omega}{2\pi}\frac{d\omega'}{2\pi} \frac{\omega A_{S+}(\omega) A_{B+}(\omega')}{(\omega+\omega')^2 + \kappa^2}.
\end{equation}
Convering to $A_S(\omega)$ gives
\begin{equation}
    F_\kappa = 2 g^2\int \frac{d\omega}{2\pi}\frac{d\omega'}{2\pi} \frac{\omega A_{S}(\omega) A_{B+}(\omega')}{(\omega+\omega')^2 + \kappa^2}\frac{1}{1-e^{-\beta_S \omega}}.
\end{equation}
Using the fact that $\omega A_S(\omega)$ is symmetric in $\omega$, we may symmetrize the remaing function of $\omega$ without changing the integral. The result is
\begin{equation}
     F_\kappa = g^2\int \frac{d\omega}{2\pi}\frac{d\omega'}{2\pi} \frac{\omega A_{S}(\omega) A_{B+}(\omega') [\omega^2+\omega'^2 +\kappa^2 - 2 \omega \omega' \coth \frac{\beta_S \omega}{2}] }{[(\omega+\omega')^2 + \kappa^2][(\omega-\omega')^2 + \kappa^2]}.
\end{equation}

Now the only potentially negative part of this expression is the function 
\begin{equation}
    f=\omega^2 + \omega'^2 +\kappa^2 - 2 \omega \omega' \coth \frac{\beta_S \omega}{2}.
\end{equation}
It is interesting to ask under what conditions $f \geq 0$. It may be written as 
\begin{equation}
    f = \left( \omega' - \omega \coth \frac{\beta_S \omega}{2} \right)^2 + \kappa^2 + \omega^2 \left( 1- \coth^2 \frac{\beta_S \omega}{2}\right). 
\end{equation}
The function $\omega^2 (1-\coth^2 (\omega/2))$ is symmetric and monotonically increasing for positive $\omega$. Its value at $\omega=0$ is $-\frac{4}{\beta_S^2}$. Hence it follows that if $\kappa$ is large enough, the function $\kappa^2 + x^2 (1- \coth^2 (x/2))$ is non-negative. From this we conclude that $F_\kappa \geq 0$ provided
\begin{equation}
    \kappa \geq \frac{2}{\beta_S}.
\end{equation}

This constraint applies for any system and bath provided that: (1) the system-bath coupling is a product of two Hermitian operators and (2) we work perturbatively in the coupling.

\printbibliography

\end{document}